\documentclass[lineo]{jfm}
\usepackage{amsmath}
\usepackage[latin1]{inputenc}
\usepackage[english]{babel}
\pdfoutput=1
\usepackage{fancyhdr}
\usepackage{babel}
\usepackage{graphicx}
\usepackage[round]{natbib}
\usepackage{newtxtext}
\usepackage{newtxmath}
\usepackage{natbib}
\usepackage{hyperref}
\hypersetup{
    colorlinks = true,
    urlcolor   = blue,
    citecolor  = black,
}

\newcommand{\RomanNumeralCaps}[1]
\linenumbers
\title{Free surface water-waves generated by instability of an exponential shear flow}

\author{Malek Abid\aff{1},
  Christian Kharif\aff{1}
  \corresp{\email{christian.kharif@centrale-marseille.fr}}
 }

\affiliation{\aff{1} {Aix-Marseille Universit\'e, Institut de Recherche sur les Ph\'enom\`nes Hors Equilibre, UMR 7342, CNRS}, {Centrale M\'editerran\'ee}, { {Marseille}, {13384}, {France}}
}

\begin{document}
\maketitle

\begin{abstract}
    The stability of an exponential current in water to infinitesimal perturbations in the presence of gravity and capillarity is investigated. Some new results on the generation of gravity-capillary waves are presented which supplement the previous works of Morland,  Saffman \& Yuen (1991) and Young \& Wolfe (2014), namely in finite depth. To consider  perturbations of much larger scales, a specific attention is paid to the stability of the exponential current only in the presence of gravity.
\end{abstract}

\begin{keywords}
Authors should not enter keywords on the manuscript, as these must be chosen by the author during the online submission process and will then be added during the typesetting process (see \href{https://www.cambridge.org/core/journals/journal-of-fluid-mechanics/information/list-of-keywords}{Keyword PDF} for the full list).  Other classifications will be added at the same time.
\end{keywords}

\section{Introduction} 
\vspace{0.1cm}
Generally, velocity profiles of currents existing in the ocean are depth dependent. Wind effect at the sea surface generates a vertically sheared current in water. Ebb and flood currents due to the tide present velocity profiles varying vertically as well as currents produced by river discharge in estuaries. \cite{Zippel2017} collected measurements of the velocity of vortical currents in the water column, at the Mouth of the Columbia River. The characteristic thickness of the vortical layer induced by wind is generally very thin. Wind induced shear currents are unstable to capillary and short gravity-capillary waves. The characteristic thickness of the shear layer due to tidal currents and river discharge is significantly larger and concerns the regime of gravity waves. The stability of shear currents due to wind is well documented whereas it is not the case for larger characteristic shear layer thickness. To the best of our knowledge the stability of oceanic shear currents to gravity wave perturbations has not yet been investigated. 
\newline
The generation of  capillary and short gravity-capillary waves on deep water due to the instability of the underlying current of depth-dependent vorticity has been investigated by many authors. Among them we can cite \cite{Stern1973}, \cite{Voronovich1980}, \cite{Caponi1991}, \cite{Morland1991}, \cite{Shrira1993}, \cite{Miles2001}, \cite{Zhang2005} and \cite{Young2014}.
\newline
\cite{Stern1973} were the first to show that the Rayleigh inflection point theorem (for rigid boundaries) is no more valid with a free surface. They used a piecewise constant vorticity profile to model the underlying current. Later, using the same profile, \cite{Caponi1991} showed that a necessary condition for unstable modes is that $u_{0s} > c_m$ where $u_{0s}$ is the surface velocity and \ $c_m = (4 g \sigma/\rho)^{1 / 4}$ is the minimum gravity-capillary wave speed for a stagnant fluid. They also showed that unstable modes then exist when the characteristic thickness of the vortical layer exceeds a critical value which depends on $u_{0s}$. Later on, \cite{Morland1991} have addressed the same problem using three distinct smooth profiles, in a fluid of
infinite depth, too. They used an exponential profile, the error function profile and the integrated error function profile. They came to the same conclusions. The transition to instability, for the smooth profiles, is an exchange of stability corresponding to the vanishing of the complex phase velocity, $c$, of the perturbations. \cite{Shrira1993}, within the framework of 3D flows derived an analytic approximate dispersion relation for linear gravity-capillary waves travelling on arbitrary underlying currents. At first order, he considered as example the stability of the exponential current in infinite depth to 2D infinitesimal gravity-capillary waves. \cite{Young2014} found, in deep water, that exponential currents are unstable to rippling perturbations due to an interaction between surface waves and a critical layer in the water. Rippling instabilities concern capillary waves of negative intrinsic phase velocity propagating against the current and whose Doppler shifted phase velocity by the surface current, $u_{0s}$, matches the current velocity at the critical depth. Note that for the exponential velocity profile the marginal curve corresponding to $c=0$ can be obtained analytically in infinite depth. \cite{Miles2001} using a variational formulation revisited the work of \cite{Morland1991} to construct an analytical description of the linear unstable modes for the exponential velocity profile. \cite{Zhang2005} considered linear gravity-capillary waves propagating at the surface of wind induced currents on finite depth. He investigated several profiles including the exponential current in deep water. \cite{NWOGU2009} investigated numerically, in deep water, the modulational instability of gravity waves travelling at the free surface of an underlying current with an exponential profile. He found that the modulational instability was enhanced in the presence of following currents.
\vspace{0.1cm}
\newline
Most of the studies on the stability of a depth varying current have considered its linear stability to capillary and gravity-capillary waves in deep water. The goal of this paper is twofold: (i) to extend to finite depth some previous results on the stability of the exponential current in the presence of surface tension and (ii) to investigate the stability of the exponential current to gravity wave perturbations.
\vspace{0.1cm}
\newline
Note that the computation of steadily propagating nonlinear water waves at the surface of a depth-dependent current requires firstly to investigate its stability.
\vspace{0.1cm}
\newline
In section 2 the equations of the linear stability problem are presented within the framework of incompressible and inviscid fluid. In section 3, the stability analysis focuses on the exponential current in water. Two examples of exponential currents measured during laboratory experiments and {\em in situ} are presented. We extend some results of \cite{Morland1991} obtained in deep water to finite depth and derived an analytic expression of the critical characteristic thickness of the shear. Section 4 is devoted to conclusion and perspective.

\section{Mathematical formulation}
\vspace{0.1cm}
We consider water waves propagating at the free surface of an inviscid and incompressible fluid governed by the following equations
\begin{eqnarray}
\nabla \cdot \bf{u}=0, \hspace{0.8cm} \label{continuity}
\\
\frac{d \bf{u}}{dt} =-\frac{\nabla p}{\rho}+\bf{g}, \label{Euler_eq} 
\end{eqnarray}
where ${\bf{u}}=(u,v)$ is the fluid velocity, $p$ is the pressure, $\rho$ is the fluid density, $\bf{g}$ is the acceleration due to gravity, $u$ and $v$ are the longitudinal and transverse components of the velocity, respectively, and
\begin{equation}
\frac{d}{dt}=\frac{\partial}{\partial t} + \bf{u} \cdot \nabla , \nonumber
\end{equation}
where $t$ is the time and $\nabla = (\partial/\partial x, \partial/\partial y)$, $x$ and $y$ are the longitudinal and vertical coordinates, respectively.
\vspace{0.1cm}
\newline
Equation (\ref{continuity}) corresponds to mass conservation whereas equation (\ref{Euler_eq}) is the Euler equation.
\vspace{0.1cm}
\newline
The boundary conditions are
\begin{equation}
v=\frac{\partial \eta}{\partial t} + u \frac{\partial \eta}{\partial x} \qquad \mathrm{on} \quad
y=\eta(x,t),
\label{kinematic_condition}
\end{equation}
where $\eta(x,t)$ is the free surface elevation,
\newline
and
\begin{equation}
p_a-p=\frac{\sigma}{R} \quad \mathrm{on} \quad y=\eta(x,t), \nonumber
\end{equation}
where $p_a$ is the atmospheric pressure at the interface and $\sigma$ is the surface tension coefficient.
The curvature is
\begin{equation}
\frac{1}{R}=\frac{\frac{\partial^2\eta}{\partial x^2}}{(1+(\frac{\partial \eta}{\partial x})^2)^{3/2}} . \nonumber
\end{equation}
The atmospheric pressure $p_a$ is set to zero without loss of generality. Consequently, the jump of pressure at the interface becomes
\begin{equation}
    p=-\frac{\sigma}{R}.
\label{Laplace_Law}
\end{equation}
\vspace{0.1cm}
In deep water 
\begin{equation}
    (u,v) \rightarrow (0,0) \quad \mathrm{as} \quad y \rightarrow -\infty.
\label{deep_water_condition}
\end{equation}
In finite depth
\begin{equation}
v=0 \quad \mathrm{on} \quad y=-h,
\label{bottom_condition}
\end{equation}
where $h$ is the depth.
\vspace{0.1cm}
\newline
Equations (\ref{kinematic_condition}), (\ref{Laplace_Law}), (\ref{deep_water_condition}) and (\ref{bottom_condition}) are the kinematic boundary condition, the Laplace law and the bottom condition, respectively.
\vspace{0.1cm}
\newline
We consider the stability of the basic steady state $(u_0(y), p_0(y))$ solution of the system of equations (\ref{continuity}) and (\ref{Euler_eq}) to small perturbations $\bf{u'}$$(x,y,t)=(u',v')$ and $p'(x,y,t)$ where $u_0(y)$ is the basic velocity profile and $p_0(y)=-\rho g y$ the pressure.
\vspace{0.1cm}
\newline
The continuity equation and linearized Euler equation read
\begin{equation}
\label{perturbation-continuity}
\frac{\partial u'}{\partial x} + \frac{\partial v'}{\partial y} =0.
\end{equation}
\begin{equation}
\label{euler-perturbation-momentum}
\left\{
\begin{array}{cc}
& \frac{\partial u'}{\partial t} + u_0 \frac{\partial u'}{\partial x}  + v' \frac{\partial u_0}{\partial y} = -\frac{1}{\rho} \frac{\partial p'}{\partial x}, \\
& \frac{\partial v'}{\partial t} + u_0 \frac{\partial v'}{\partial x} = -\frac{1}{\rho} \frac{\partial p'}{\partial y}. 
\end{array} \right.
\end{equation}
\vspace{0.1cm}
\newline
The linearized boundary conditions read
\begin{equation}
    \frac{\partial \eta'}{\partial t} +u_{0s} \frac{\partial \eta'}{\partial x}=v' \quad \mathrm{on} \quad y=0,
\label{linearized-kinematic-boundary-condition}
\end{equation}
where $u_{0s}=u_0(0)$,
\newline
and
\begin{equation}
    p'=\rho g \eta'-\sigma \frac{\partial^2 \eta'}{\partial x^2}  \quad \mathrm{on} \quad y=0.
\label{linearized-dynamic-boundary-condition}
\end{equation}
In deep water 
\begin{equation}
(u',v') \rightarrow (0,0) \quad \mathrm{as} \quad y \rightarrow -\infty,
\label{infinite-depth-condition}
\end{equation}
and in finite depth
\begin{equation}
v'(-h)=0.
\label{bottom-condition-perturbation}
\end{equation}
The solution of the linearized problem is sought in the following form
\begin{equation}
(u',v',p',\eta')=(u_1(y),v_1(y),p_1(y),\eta_1)\exp(ik(x-ct)),
\label{perturbation_form}
\end{equation}
where $k$ is the perturbation wavenumber and $c$ its complex phase velocity. 
\newline
The system of equations (\ref{perturbation-continuity})-(\ref{bottom-condition-perturbation}) reduces the the Rayleigh equation
\begin{equation}
    \frac{d^2 v_1}{dy^2}-(k^2 + \frac{\frac{d^2 V}{dy^2}}{V})v_1=0,
\label{Rayleigh equation}
\end{equation}
with the following boundary conditions
\begin{equation}
   V^2(0)\frac{dv_1}{dy}(0)-(\frac{dV}{dy}(0)V(0)+g+\frac{\sigma k^2}{\rho})v_1(0)=0,
\label{perturbation-kinematic+dynamic-boundary-conditions}
\end{equation}
where $V(y)=u_{0}-c$ 
\newline
\begin{equation}
v_1 \rightarrow 0 \quad \mathrm{as} \quad y \rightarrow -\infty, 
\label{bottom_condition_deep}
\end{equation}
in deep water and,
\begin{equation}
v_1(-h)=0,
\label{bottom_condition_finite_depth}
\end{equation}
in finite depth.
\section{Linear stability of the exponential current}
\begin{figure}
    \centering
    \includegraphics[width=10cm]{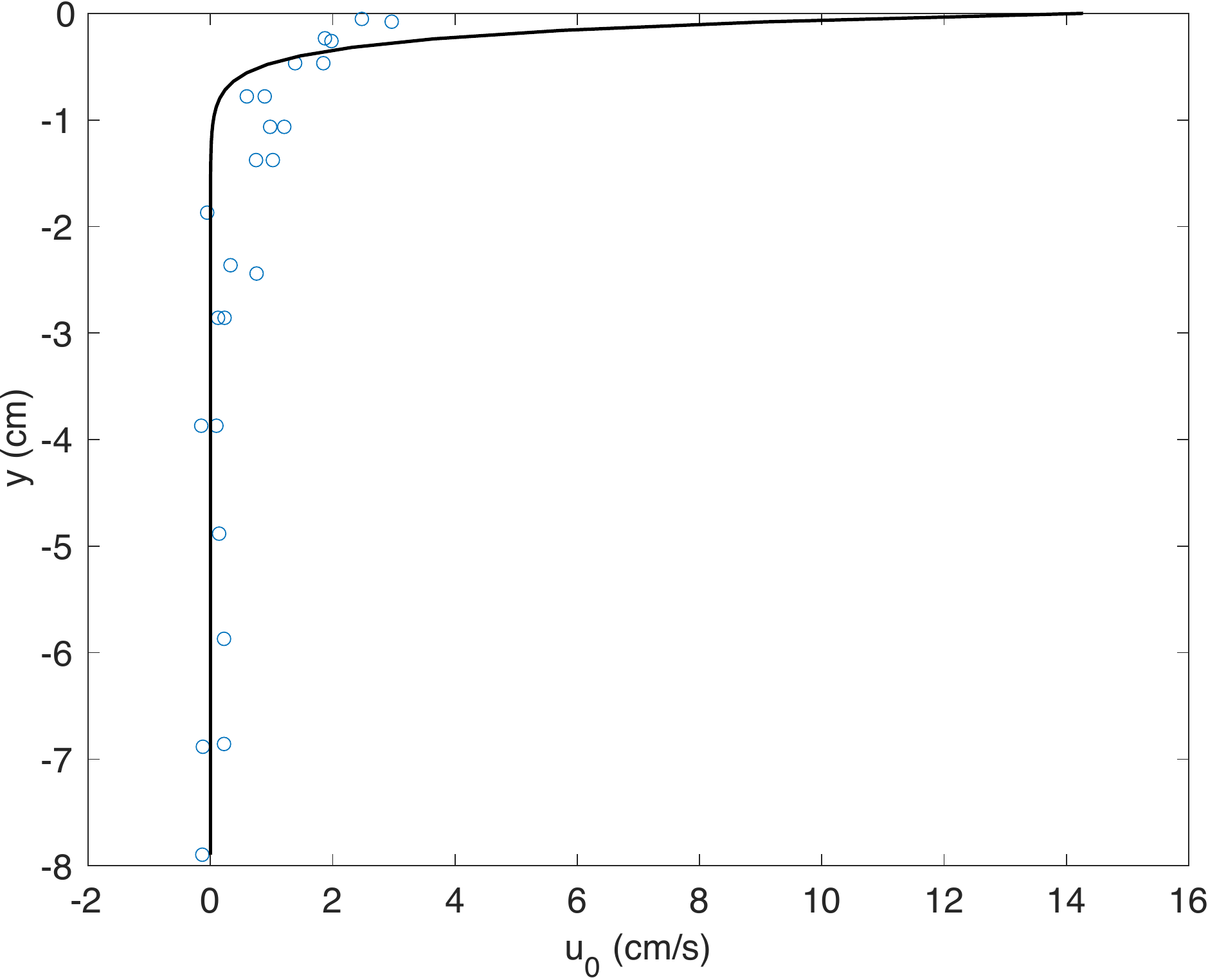}
    \caption{Current velocity profile in water. Crosses: experimental data obtained in the wind wave facility IRPHE/Pytheas (H. Branger, private communication). The solid line is the exponential profile (equation (\ref{eq_prof_exp})) that fits the experimental data obtained with a wind velocity (extrapolated to the standard altitude of $10\ m$) of $6\ m/s$ and a water depth of $20\ cm$.}
    \label{fig:fir_exp}
\end{figure}

\begin{figure}
    \centering
    \includegraphics[width=0.48\linewidth]{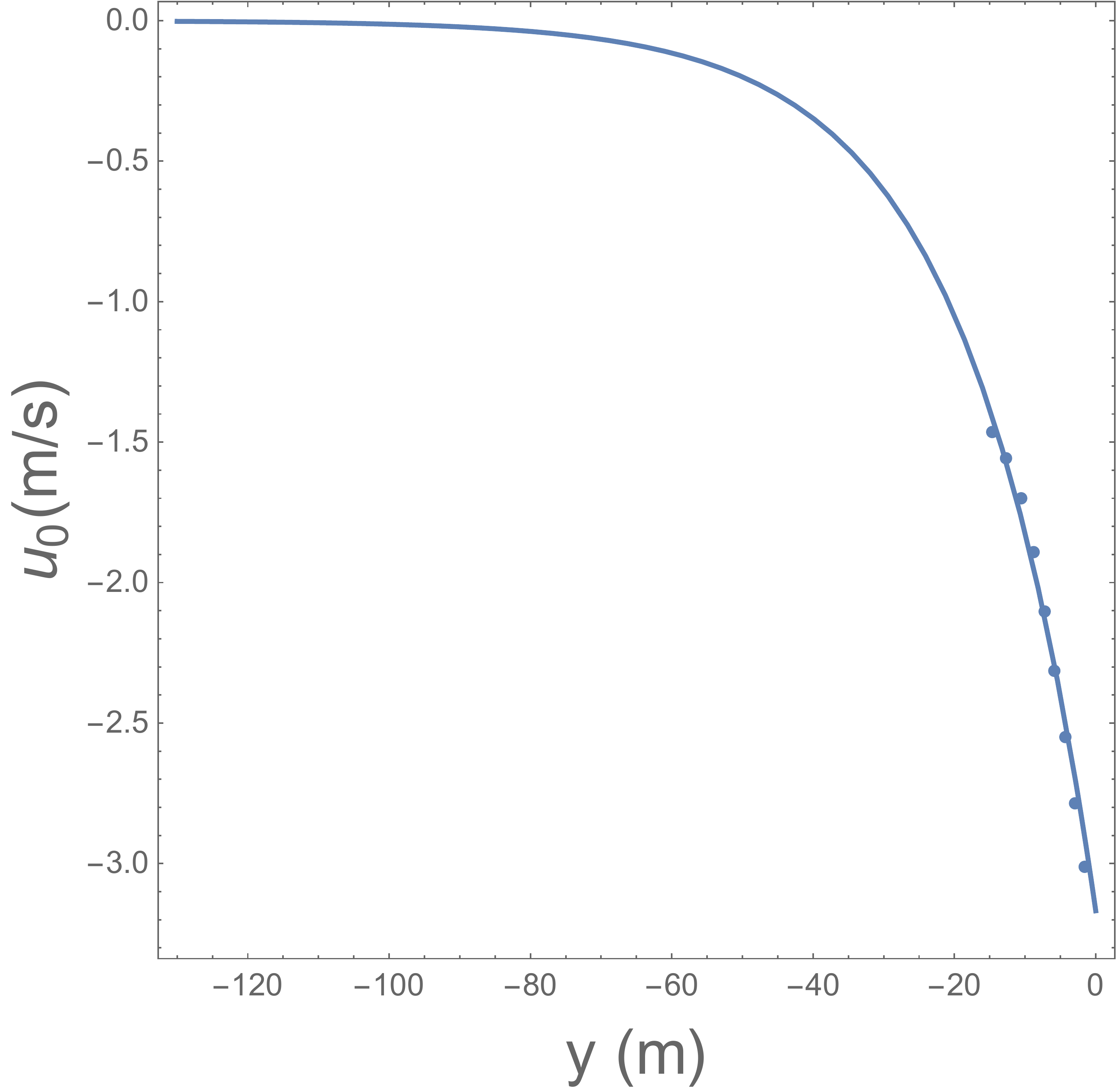}
    \includegraphics[width=0.47\linewidth]{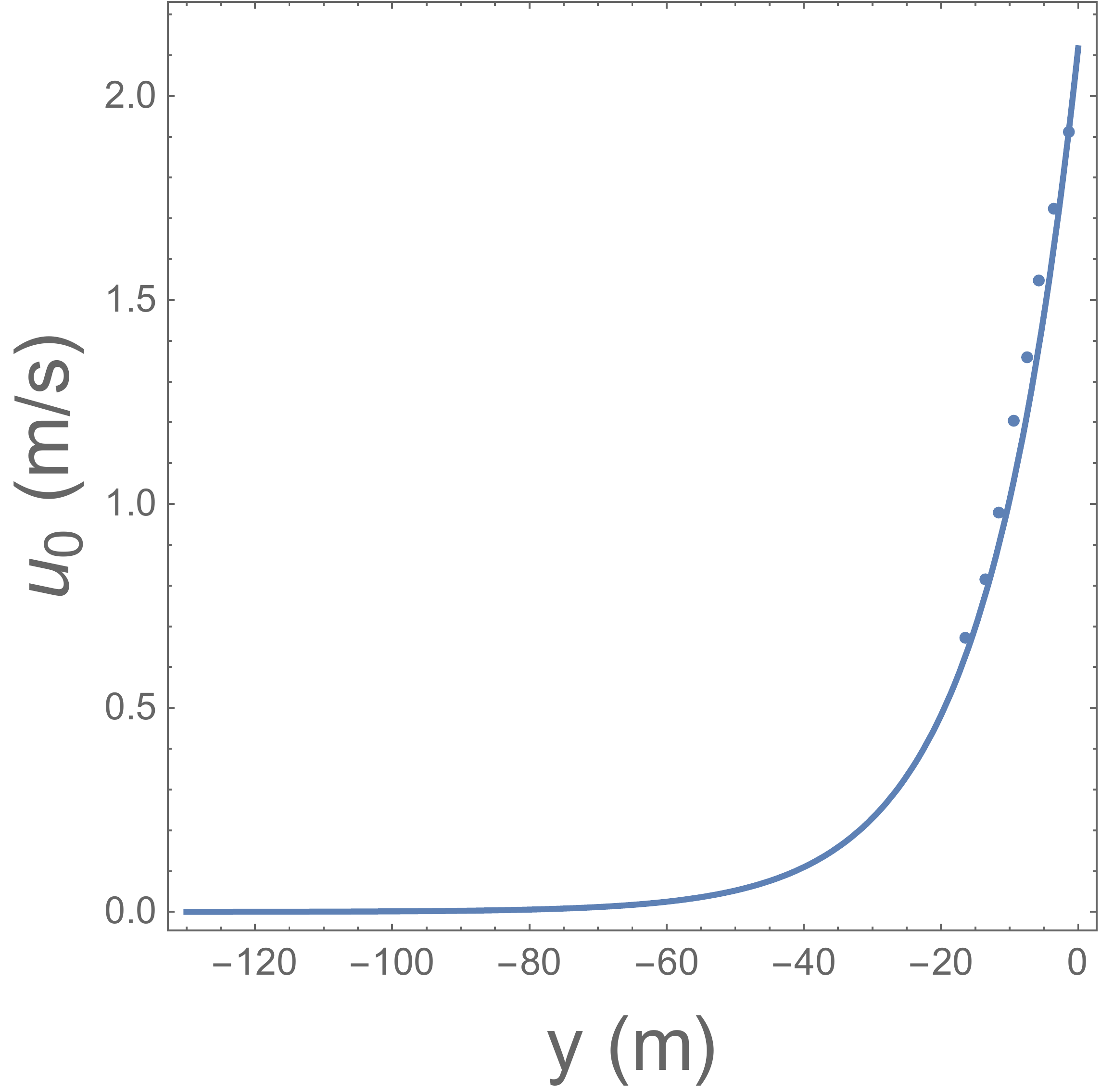}
    \caption{Ebb and flood velocity profiles measured at the Mouth of the Columbia River (dots). The solid lines are the exponential profiles given by equation (\ref{eq_prof_exp}) that fit the current profiles measured in-situ by \cite{Zippel2017}. Ebb velocity profile (left) and flood velocity profile (right).}
    \label{Ebb_Flood_Profiles}
\end{figure}

\vspace{0.1cm}
Wind effect at the sea surface is twofold: it generates a vertically sheared current in water and then short waves. In figure \ref{fig:fir_exp} is plotted the exponential profile that fits the data of experiments conducted in the wind wave facility of IRPHE/Pytheas at Luminy (Marseille). Regular monochromatic linear waves are mechanically generated under the action of wind. Owing to the weakness of the wave steepness of the surface waves ($ak=0.055$), the underlying current is mainly due to the wind. The following form approximates the experimental induced current in water 
\begin{equation}
    u_0(y) = u_{0s} \exp (\delta y), \quad -h < y < 0.
    \label{eq_prof_exp}
\end{equation}
where $1/\delta$ is the characteristic thickness of the shear layer. Note that $u_{os} < c_m$ and consequently the exponential current is stable.
\newline
In addition to the experimental profile measured in the wind wave facility, two tidal currents measured {\em in-situ} by \cite{Zippel2017} at the Mouth of the Columbia River are shown in figure \ref{Ebb_Flood_Profiles}.
\newline
Currents, in the upper ocean, of exponential profile type may exist at very different spatial scales ranging from few millimeters to several ten meters. The stability of the exponential current to infinitesimal capillary waves and gravity-capillary waves is well documented whereas it is not the case for pure gravity waves.
\vspace{0.10cm}
\newline
Within the framework of the exponential velocity profile the marginal curve corresponding to $c=0$ can be obtained analytically in infinite depth. The works previously cited concern studies where the current velocity varies between $u_{0s}$ and zero. In finite depth the exponential current varies between $u_{0s}$ and $u_{bottom}$ which does not vanish. Due to the semicircle theorem of Howard $c$ must lie in the semicircle for unstable waves and consequently cannot vanish. The curve separating stable and unstable domains can be only determined numerically. More generally, in finite depth the term $\frac{d^2V}{dy^2}/V$ of equation (\ref{Rayleigh equation}) depends on $y$ and consequently the marginal curve cannot be determined analytically.
\subsection{The Rayleigh equation and its analytical solution}
Within the framework of the exponential velocity profile in deep water, the Rayleigh equation can be integrated analytically. Miles, in Appendix A of the paper by \cite{Morland1993}, gave the exact expression of the solution in terms of the hypergeometric function $F(a,b;\alpha;\beta(y))$. Consequently, in deep water equation (\ref{Rayleigh equation}) admits the following solution
\begin{equation}
    v_1(y)=\exp(ky)F(a,b;\alpha;\beta(y)),
    \label{hypergeometric_function}
\end{equation}
where $a=K-\sqrt{1+K^2}$, $b=K+\sqrt{1+K^2}$, $\alpha=1+2K$, $\beta(y)=u_{0s} \exp(\delta y)/c$ and $K=k/\delta$.
\newline
The derivative is
\begin{equation}
    \frac{dv_1}{dy}=k\exp(ky)F(a,b;\alpha;\beta(y))-\frac{\exp(k y)}{\alpha}F(a+1,b+1; \alpha +1; \beta(y))\frac{\delta u_{0s}}{c}\exp(\delta y).
    \label{derivative_hypergeometric_function}
\end{equation}
Equation (\ref{perturbation-kinematic+dynamic-boundary-conditions}) is rewritten as follows
\begin{equation}
    (u_{0s}-c)^2 \frac{dv_1}{dy}(0)-(\delta u_{0s}(u_{0s}-c)+g+\frac{\sigma k^2}{\rho})v_1(0)=0,
    \label{dispersion_relation_on_exponential_current}
\end{equation}
with
\begin{equation}
 \frac{dv_1}{dy}(0)=k F(a,b;\alpha;u_{0s}/c) -\frac{\delta u_{0s}}{\alpha c} F(a+1,b+1; \alpha+1; u_{0s}/c) 
 \label{derivative_v1_zero}
\end{equation}
and 
\begin{equation}
 v_1(0)=F(a,b;\alpha; u_{0s}/c).
 \label{v1_zero}
\end{equation}
The complex phase velocity $c=c_r+ic_i$ is obtained by solving numerically, with the help of {\em Mathematica}, equation (\ref{dispersion_relation_on_exponential_current}) with $v_{1{y}}(0)$ and $v_1(0)$ given by equations (\ref{derivative_v1_zero}) and (\ref{v1_zero}). Note that equation (\ref{dispersion_relation_on_exponential_current}) is the dispersion relation of free surface waves travelling on an exponential current in deep water.
\vspace{0.15cm}
\newline
With $u_0$ given by (\ref{eq_prof_exp}) and $c=0$, the Rayleigh equation admits the following solution
\begin{equation}
    v_1(y)=\exp(\sqrt{k^2+\delta^2}y) .
\nonumber
\end{equation}
The boundary condition (\ref{perturbation-kinematic+dynamic-boundary-conditions}) with $c=0$ becomes
\begin{equation}
    u_{0s}^2(\sqrt{k^2+\delta^2}-\delta)-g-\frac{\sigma k^2}{\rho}=0 .
\nonumber
\end{equation}
Introducing the intrinsic phase velocity of linear gravity-capillary waves on deep water $c_0^2=g/k+\sigma k/\rho$ we obtain
\begin{equation}
    \sqrt{k^2+\delta^2}-\delta=kc_0^2/u_{0s}^2 ,
    \nonumber
\end{equation}
\begin{equation}
    K=2\frac{c_0^2/u_{0s}^2}{1-c_{0}^4/u_{0s}^4} .
    \label{analytic_neutral_curve}
\end{equation}
\begin{figure}
    \centering
    \includegraphics[width=9cm]{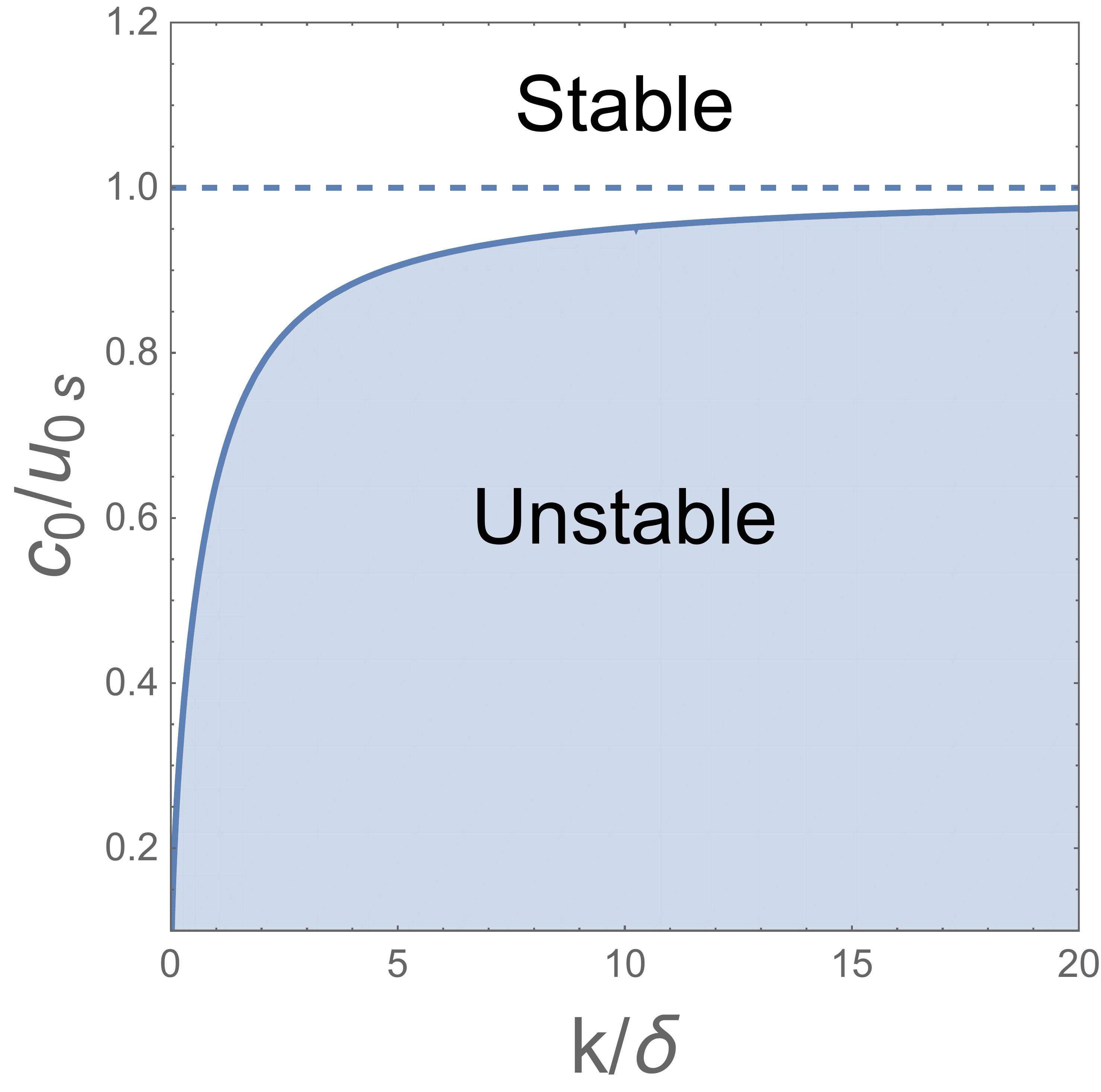}
    \caption{Stability diagram of surface waves on deep water in the plane ($c_0/u_{0s}, k/\delta$) where $c_0$ is the intrinsic phase velocity, $u_{0s}$ the surface current, $k$ the wavenumber of the perturbation and $\delta$ the inverse of the characteristic thickness of the shear. The solid line is the marginal curve and the dashed line the asymptote when $k/\delta$ goes to $\infty$}
    \label{fig:marginal_inf_dep}
\end{figure}
\begin{figure}
      \centering
      \includegraphics[width=9cm]{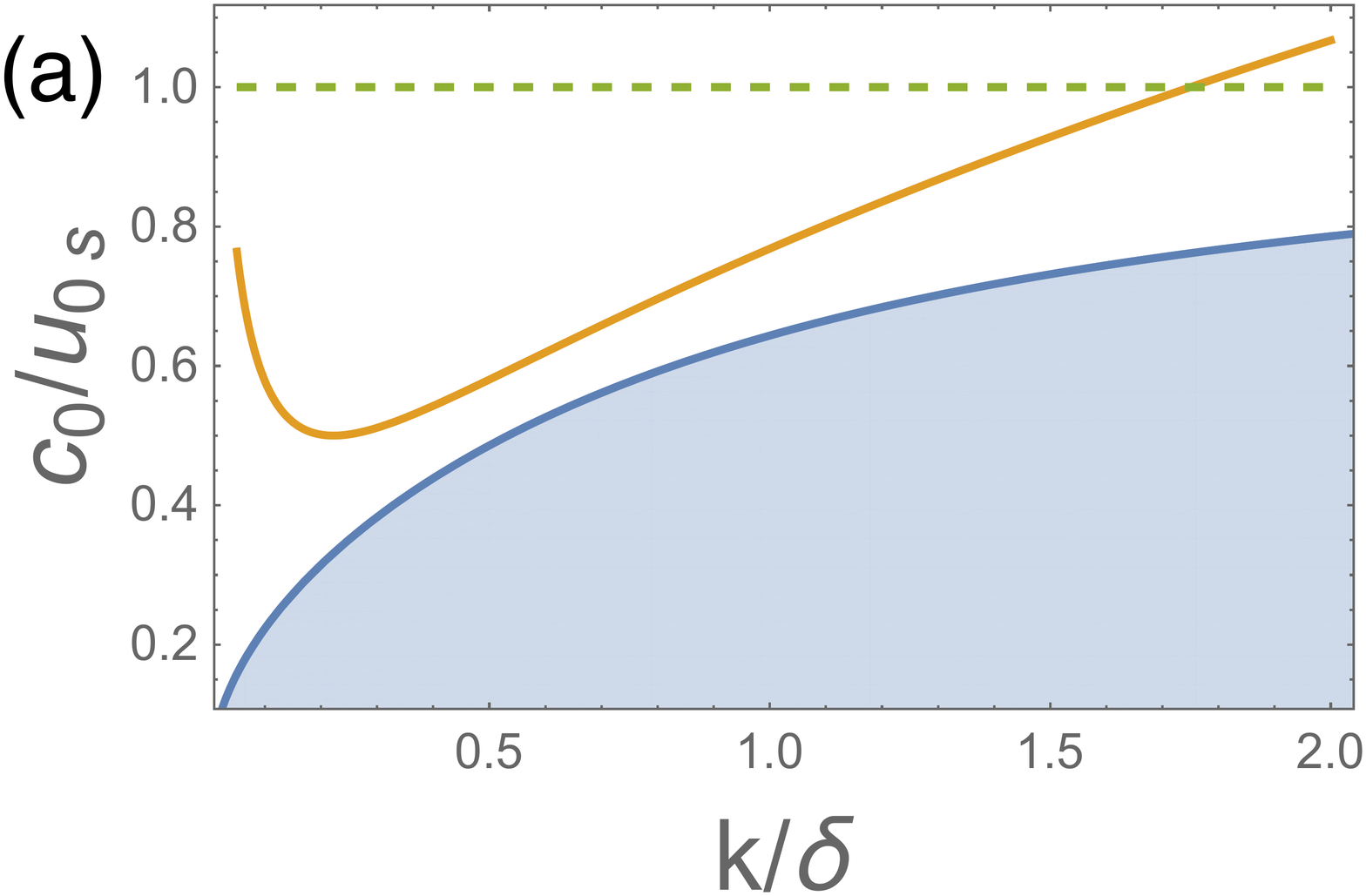}
      \includegraphics[width=9cm]{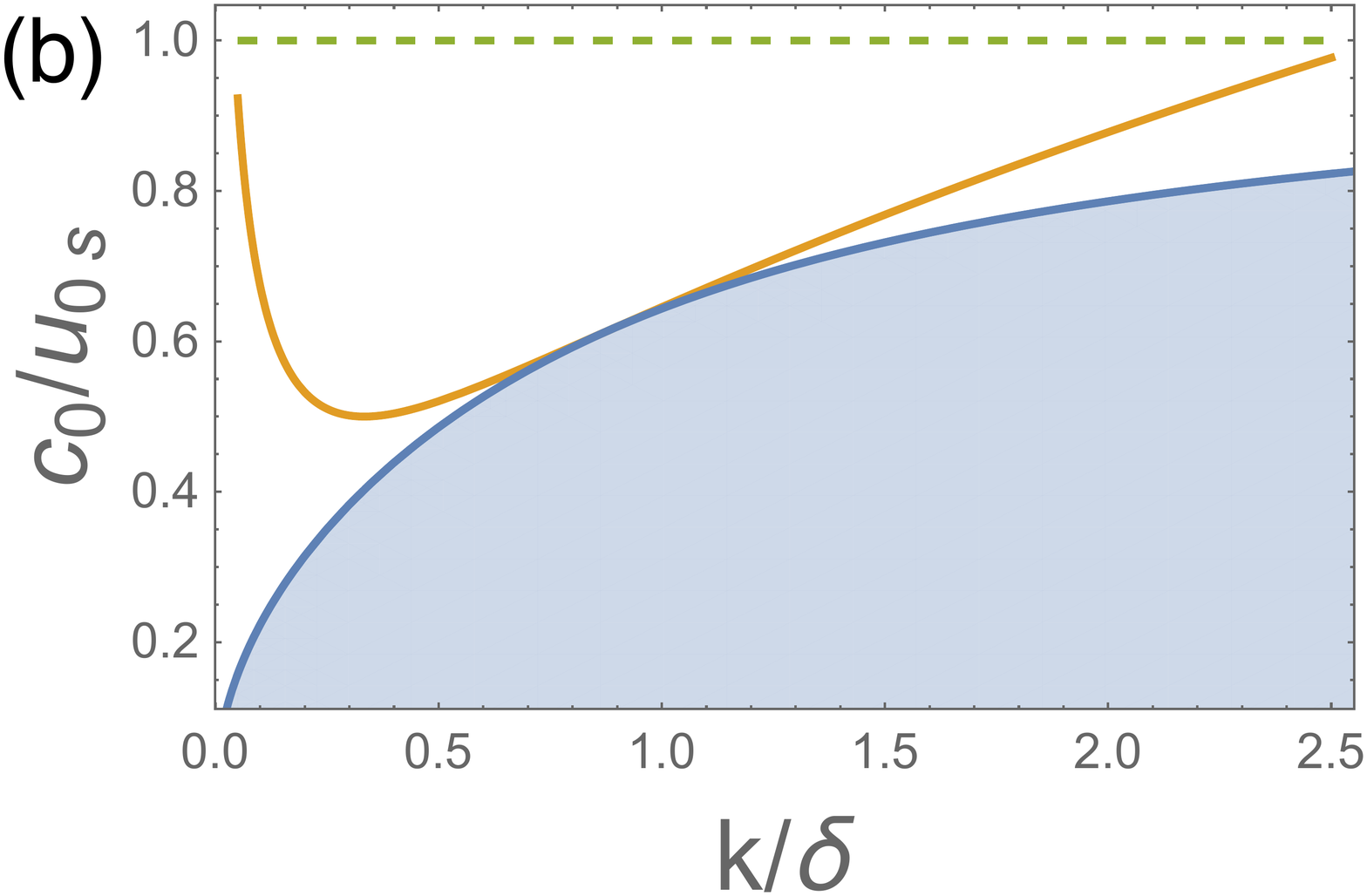}
      \includegraphics[width=9cm]{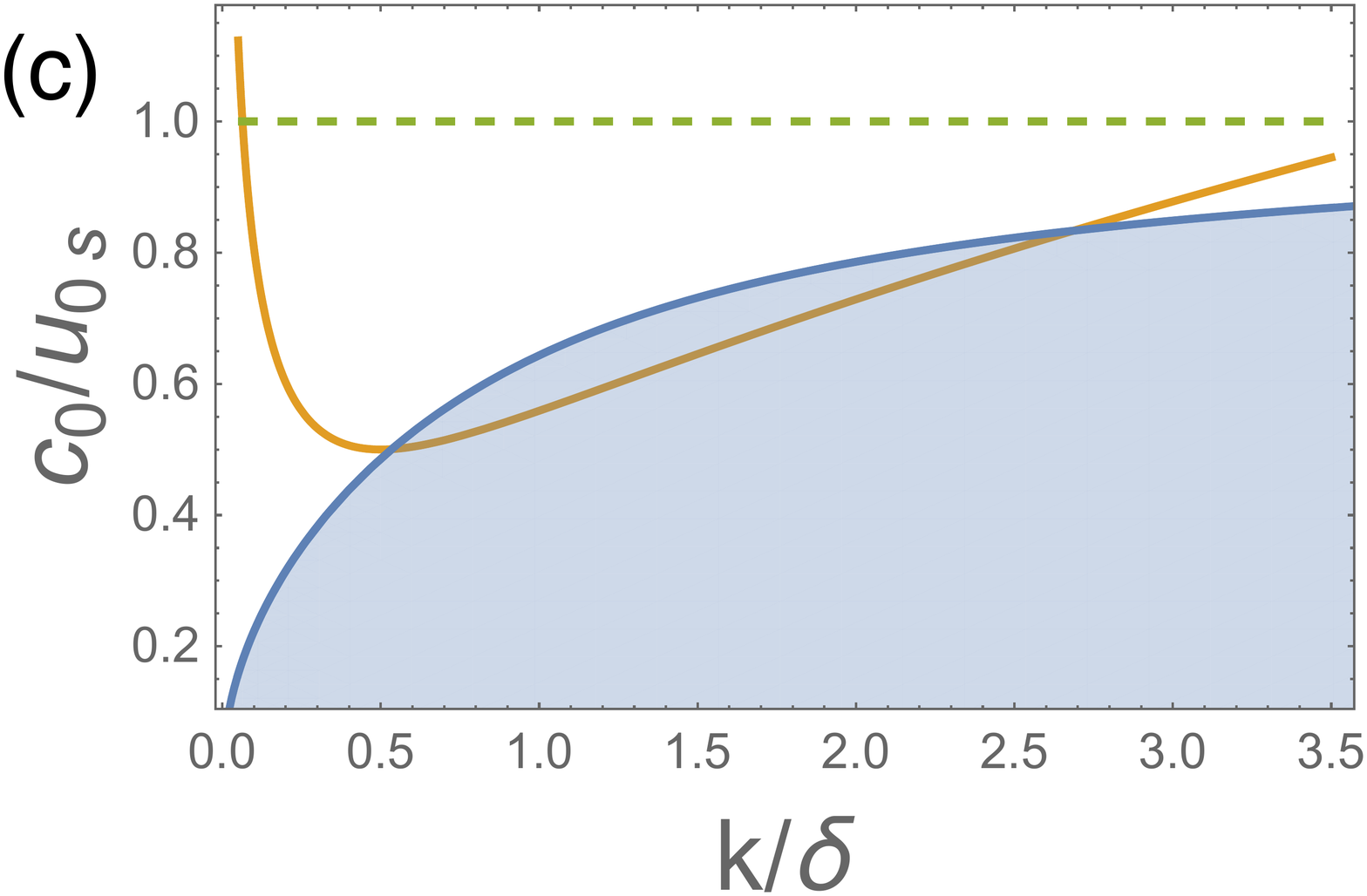}
    \caption{Color on line. Effect of the characteristic thickness, $1/\delta$, of the shear on the occurrence of instability in deep water with $u_{0s}=2\, c_m$ and $\delta=4.5\ k_m$ (a), $\delta=3\, k_m$ (b), $\delta=2\, k_m$ (c). $k_m=(\rho \,g/\sigma)^{1/2}$ is the wavenumber corresponding to the minimum of the phase velocity $c_m$. The blue solid line is the marginal curve and the orange solid line is the graph of the dimensionless linear dispersion relation of gravity-capillary waves $c_0/u_{0s}$.}
    \label{fig:marginal_inf_dep_crit_del}
\end{figure}
Equation (\ref{analytic_neutral_curve}) is the analytic expression of the marginal curve in the ($K, c_0/u_{0s}$) plane plotted in figure \ref{fig:marginal_inf_dep}. Note that equation
(\ref{analytic_neutral_curve}) applies to gravity-capillary waves as well as gravity waves.

\subsubsection{Stability of the exponential current to gravity-capillary wave perturbations}
\begin{figure}
    \centering
    \includegraphics[width=0.8\linewidth]{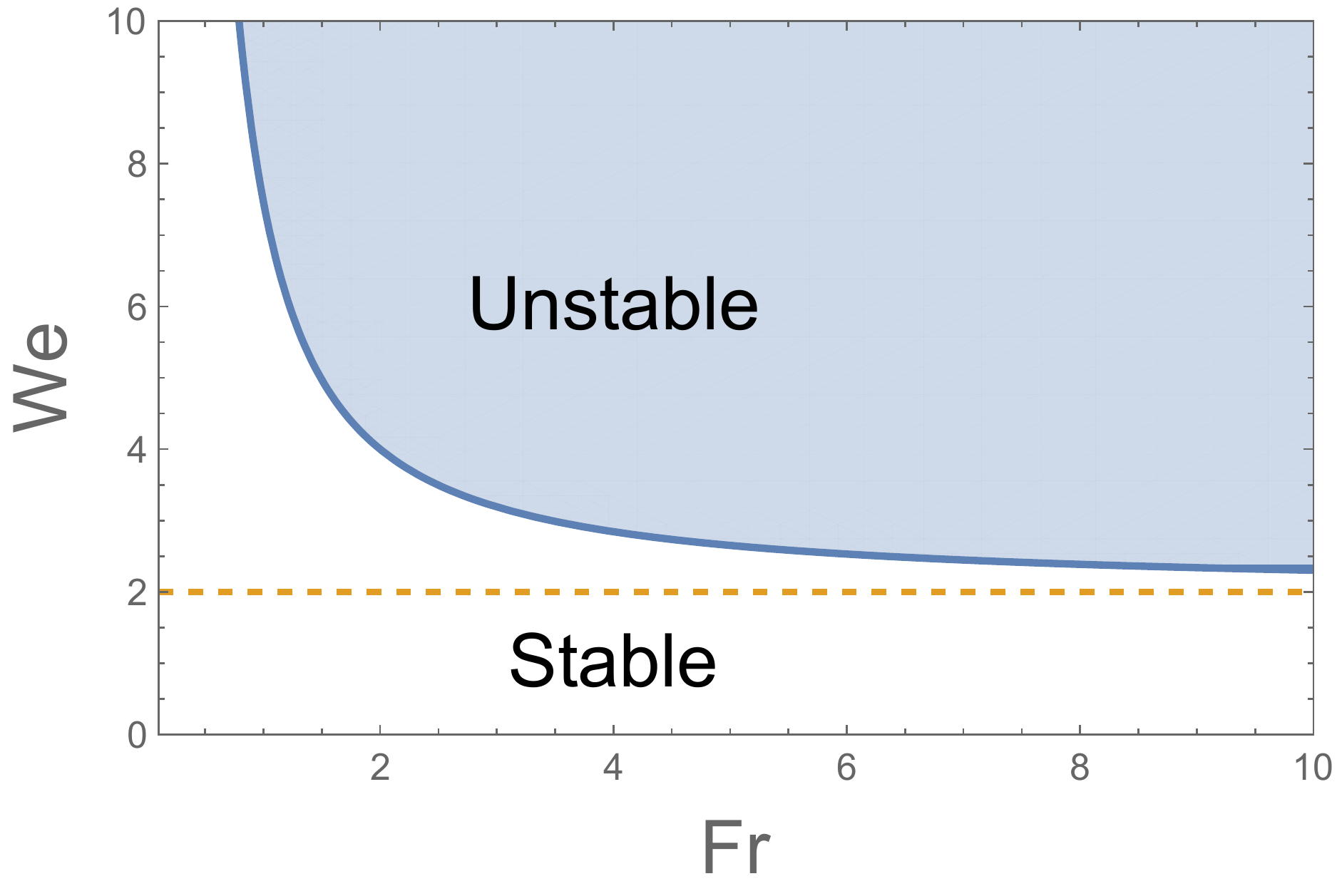}
    \caption{Stability diagram of gravity-capillary waves on deep water in the plane ($Fr$, $We$) with the Froude number $Fr=\sqrt{\delta u_{0s}^2/g}$ and the Weber number $We=\rho u_{0s}^2/(\delta \sigma)$.}
   \label{fig:Marginal_inf_depth}
\end{figure}

\begin{figure}
   \centering
   \includegraphics[width=0.8\linewidth]{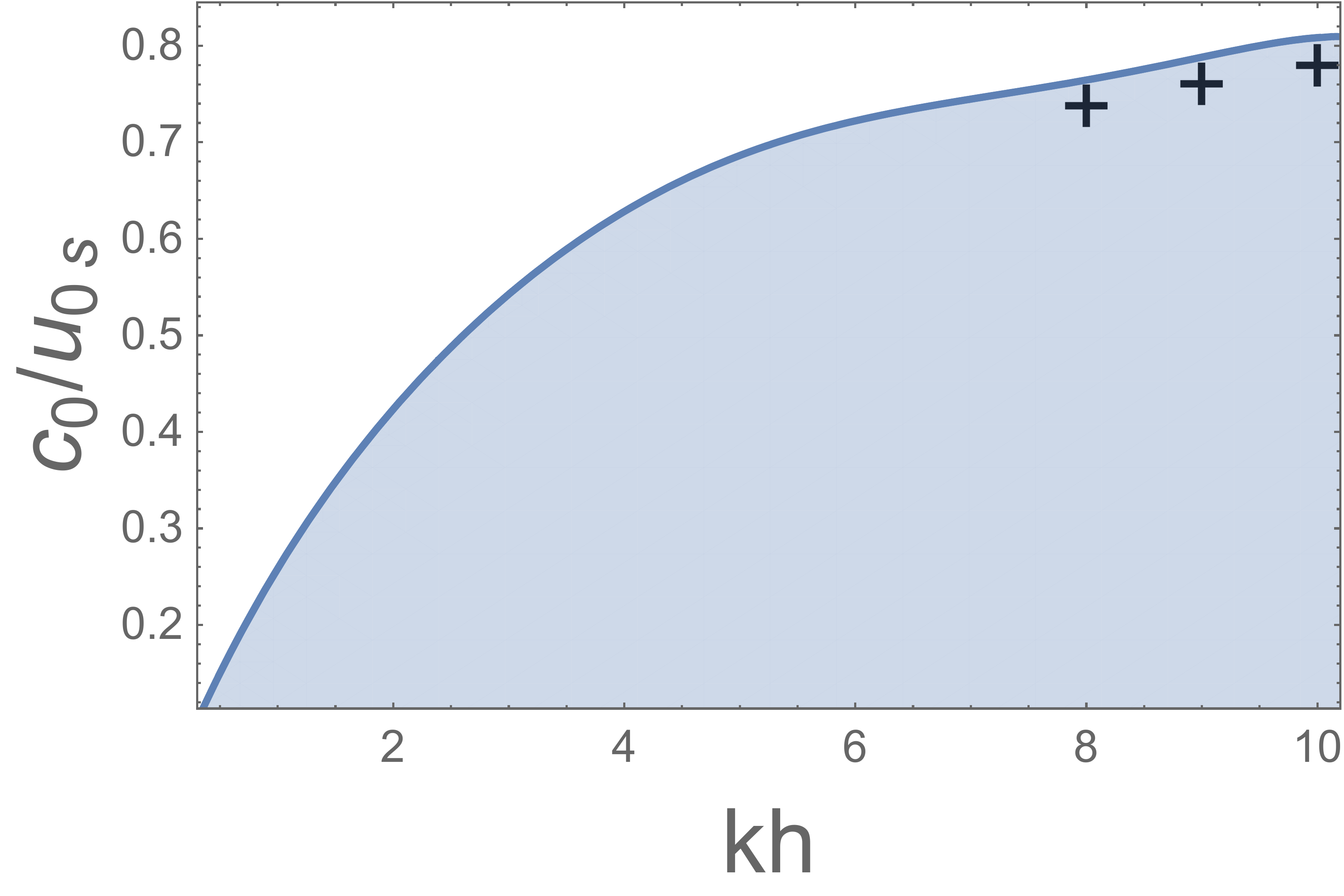}
   \includegraphics[width=0.8\linewidth]{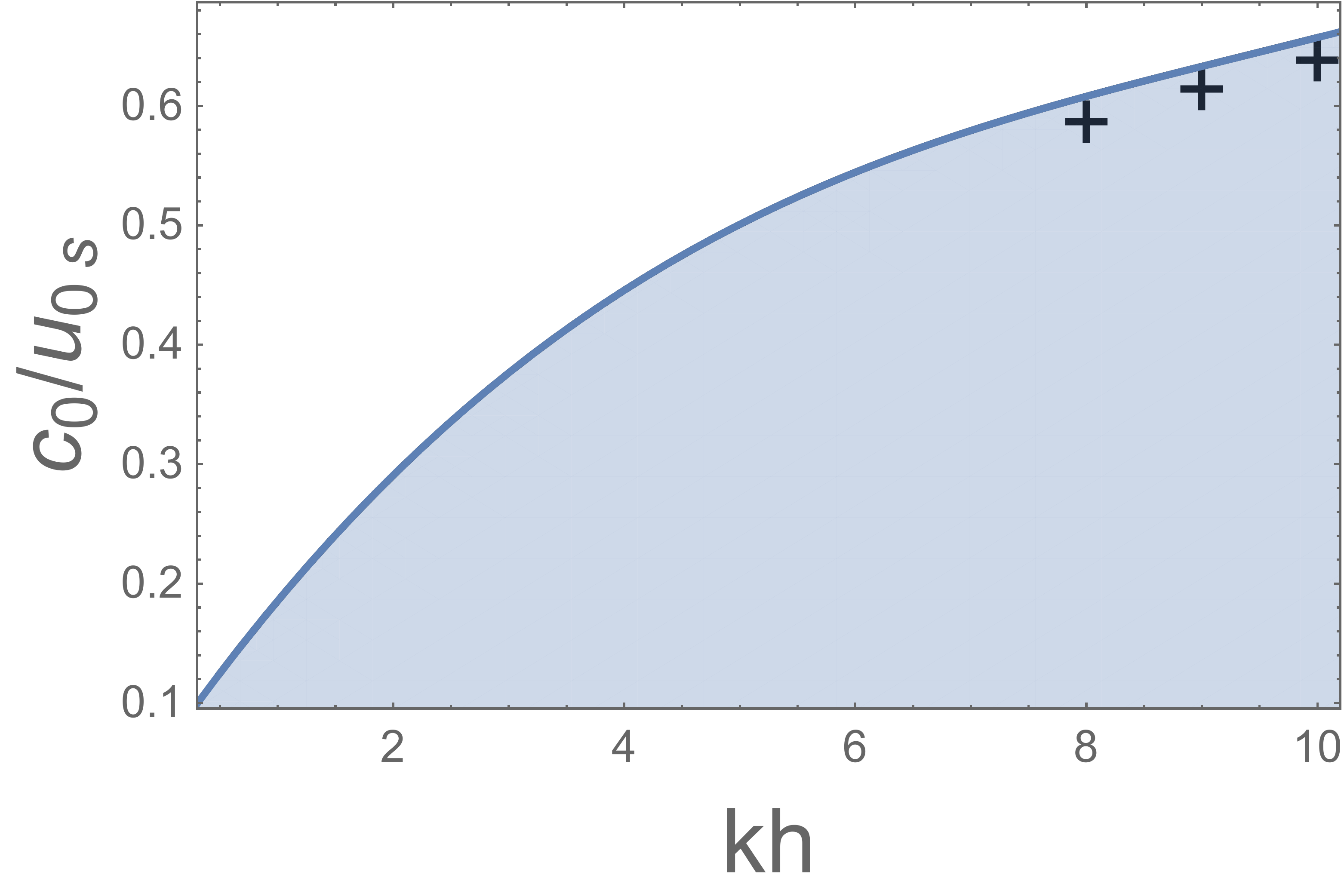}
    \caption{Stability diagram of gravity capillary waves on finite depth in the plane ($c_0/u_{0s}, kh$) where the depth $h=0.20 \, m$. $\delta h=5$ (top) and $\delta h=10$ (bottom). Crosses correspond to deep water.}
    \label{fig:marginal_fin_dep}
\end{figure}

The purpose of this subsection is not to develop a detailed stability analysis of the exponential current to gravity-capillary wave perturbations which has been done by \cite{Morland1991} and \cite{Young2014} in infinite depth, but rather to complete their investigations with some new results in infinite depth and finite depth. 
\vspace{0.1cm}
\newline
Figure \ref{fig:marginal_inf_dep_crit_del} shows the effect of the characteristic thickness $1/\delta$, in deep water, on the stability of the underlying current when the surface current satisfies the necessary condition of instability, $u_{0s}>c_m$. Three values of $1/\delta$ are introduced corresponding to stability (figure \ref{fig:marginal_inf_dep_crit_del}-a, $1/\delta=0.035 \lambda_m$, $\lambda_m=2\pi/k_m$, $k_m=\sqrt{\rho g/\sigma}$), marginal stability (figure \ref{fig:marginal_inf_dep_crit_del}-b, $1/\delta_c=0.053 \lambda_m$) and instability (figure \ref{fig:marginal_inf_dep_crit_del}-c, $1/\delta=0.08 \lambda_m$), respectively. The values of $K^+$ and $K^-$ given by equation (\ref{dimensionless_wavenumber_K+_K-}) with $u_{0s}=2c_m$ and $\delta=2k_m$, are close to $2.8$ and $0.5$, respectively. To each value of $u_{0s}$ is associated a critical characteristic thickness of the shear $1/\delta_c$ corresponding to the onset of instability. When $\delta=\delta_c$ the orange solid line is tangent to the blue solid line. Consequently, rippling instability conditions are $u_{0s}>c_m$ and $\delta<\delta_c$. The analytic expression of $\delta_c$ is
\begin{equation}
    \delta_c=\frac{u_{0s}^2-c_m^2}{2\sqrt{\sigma/\rho}}.
    \label{delta_critique}
\end{equation}
The critical characteristic shear thickness corresponding to the occurrence of rippling instability in water is $\mathcal{O}[(u_{0s}^2-c_m^2)10^{-3}\,m^{-1}]$. Note that instability condition $u_{0s}>c_m$ is involved in equation (\ref{delta_critique}), implicitly. The critical characteristic thickness of the shear layer decreases as the surface velocity $u_{0s}$ increases.
\vspace{0.1cm}
\newline
Introducing the Froude number, $Fr$, and the Weber number, $We$, we obtain the following dimensionless form of equation (\ref{analytic_neutral_curve})
\begin{equation}
    \sqrt{1+K^2}-1-\frac{1}{Fr^2}-\frac{K^2}{We}=0,
\label{dimesionless_marginal_equation}
\end{equation}
with $Fr=\sqrt{\delta u_{0s}^2/g}$, $We=\rho u_{0s}^2/(\delta \sigma)$.
\vspace{0.10cm}
\newline
The positive roots of equation (\ref{dimesionless_marginal_equation}) are
\begin{equation}
K^{\pm}=We\sqrt{\frac{1}{2}-\frac{1}{We}-\frac{1}{We Fr^2}\pm\sqrt{(\frac{1}{2}-\frac{1}{We})^2-
\frac{1}{We Fr^2}}}.
\label{dimensionless_wavenumber_K+_K-}
\end{equation}
These roots correspond to the intersection points between the marginal and dispersion curves as shown in figure \ref{fig:marginal_inf_dep_crit_del}-c. Instability occurs when $K^-<K<K^+$. When these intersection points merge, as shown in figure \ref{fig:marginal_inf_dep_crit_del}-b, the marginal and dispersion curves are tangent and $K^+=K^-=K_{marginal}=\sqrt{W_e^2/4-1}$.
\vspace{0.1cm}
\newline
The marginal curve in the ($Fr$, $We$) plane is obtained by introducing the expression of the marginal wavenumber in equation (\ref{dimesionless_marginal_equation})
\begin{equation}
    We=\frac{2}{Fr^2}(1+Fr^2+\sqrt{1+2Fr^2}) .
\label{marginal_equation_We-Fr}
\end{equation}
Note that the necessary and sufficient condition for $K^+$ and $K^-$ to be real is
\begin{equation}
    We \geq \frac{2}{Fr^2}(1+Fr^2+\sqrt{1+2Fr^2}).
 \end{equation}
In figure \ref{fig:Marginal_inf_depth} is plotted the marginal curve in the ($Fr$, $We$) plane. In infinite depth the velocity profile is stable $\forall \, Fr$ when 
\begin{equation}
    We < 2 .
    \label{weber_critique}
\end{equation}
{\bf{Remark}}: Within the framework of the stability of a thin layer of inviscid fluid having a linear velocity profile, \cite{Miles1960} demonstrated that $We=\rho u_{0s}^2 h/\sigma < 3$, where $h$ is the depth of the layer, is a sufficient condition for stability. Note that the condition (\ref{weber_critique}) has been obtained under different conditions. In Appendix B we have considered the stability of a thin film of liquid in an exponential shearing flow. We found that (i) the dimensionless growth rate increases as the depth decreases and (ii) the bandwidth of the characteristic shear layer thickness corresponding to instability decreases as the depth decreases.
\vspace{0.1cm}
\newline
As mentioned previously, in finite depth the marginal curve cannot be obtained analytically. In Appendix A is presented the numerical approach to determine the marginal curve and the complex phase velocity, $c$, of the perturbations. Figure \ref{fig:marginal_fin_dep} shows two marginal curves in the plane ($c_0/u_{0s}, kh$) for two values of $\delta h$ where $h$ is the depth. The stable domain is located above the marginal curve. The size of the stable domain increases as the characteristic thickness of the shear decreases for fixed values of the depth.
\vspace{0.10cm}
\newline
To summarize, our contributions to the problem of the linear stability of the exponential current to gravity-capillary waves are (i) the derivation of an analytic expression of the critical characteristic thickness of the shear (equation (\ref{delta_critique})) (ii) a stability criterion in deep water when the Weber number $We$ is less than $2$ independently of the Froude number $Fr$ (iii) the plots of stability diagrams in finite depth and (iv) the stability of a thin film of liquid in an exponential shearing flow.

\subsubsection{Gravity wave instabilities}
\begin{figure}
    \centering
    \includegraphics{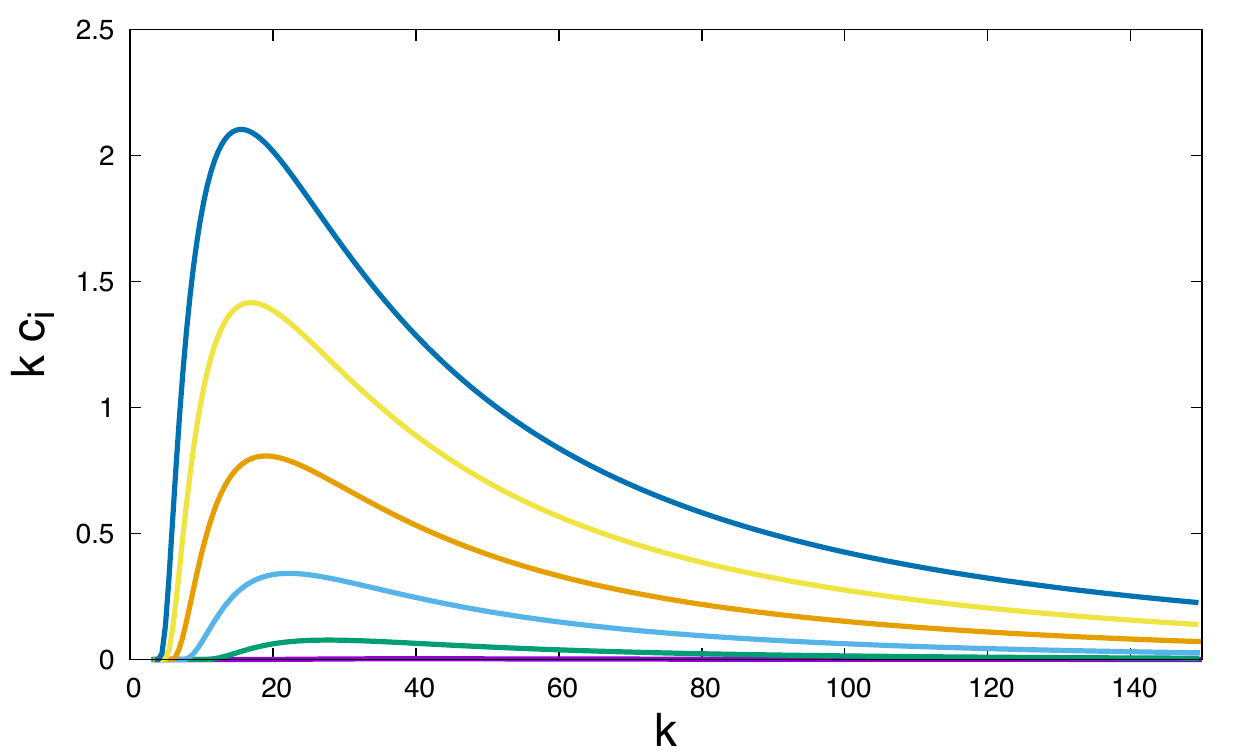}
    \includegraphics{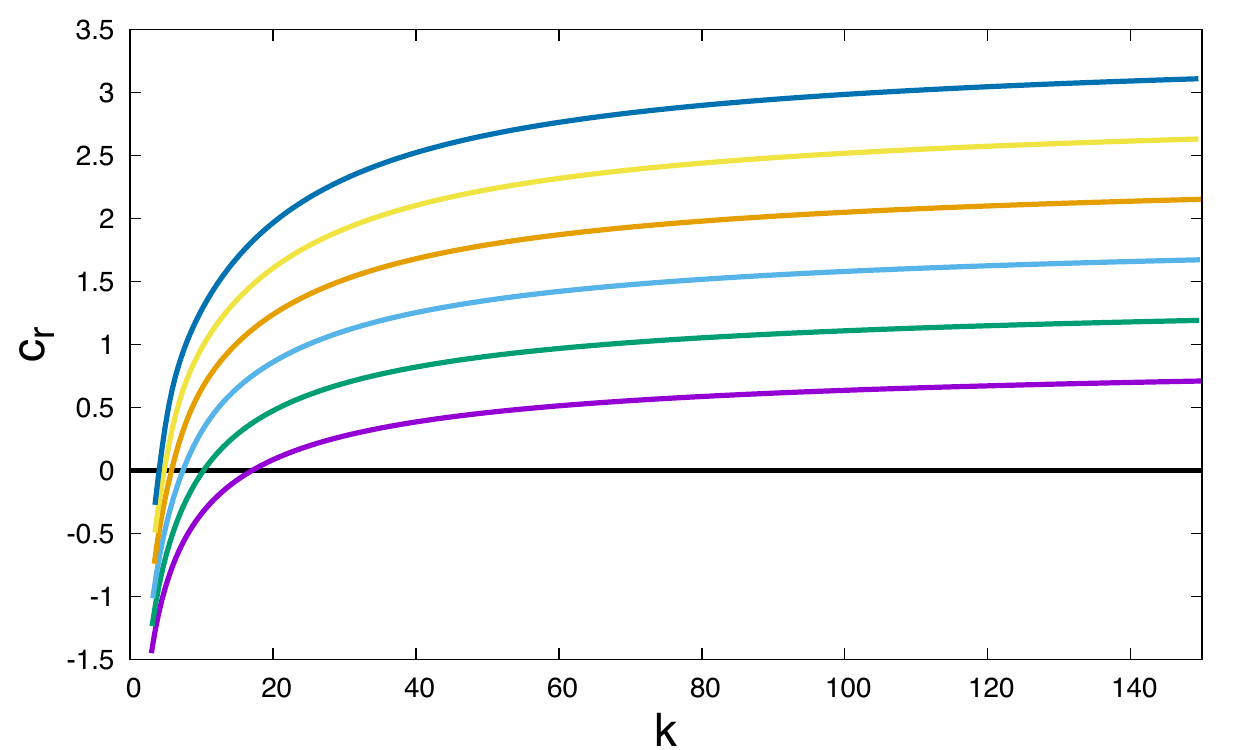}
    \caption{Color on line. Dimensional growth rates of gravity waves (top) and phase velocities (bottom), against the wavenumber. For $\delta=10\ m^{-1}$ and $u_{0s}$ increases from $1\ m.s^{-1}$ to $3.5\ m.s^{-1}$ by increment of $0.5\ m.s^{-1}$ (the Froude number varies from 1 to 3.5). The lower curves (in the top and bottom figures) correspond to $u_{0s}=1\ m.s^{-1}$.}
    \label{fig:kcicr_delta_10}
\end{figure}

\begin{figure}
    \centering
    \includegraphics{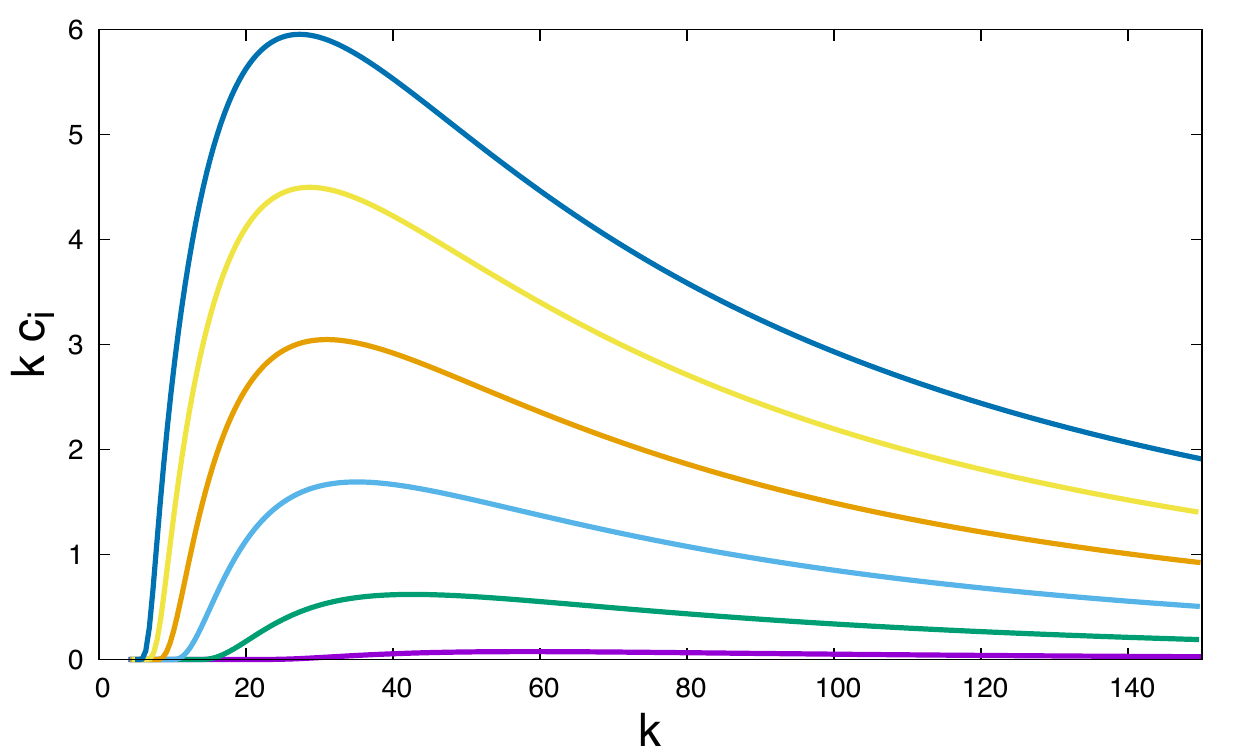}
    \includegraphics{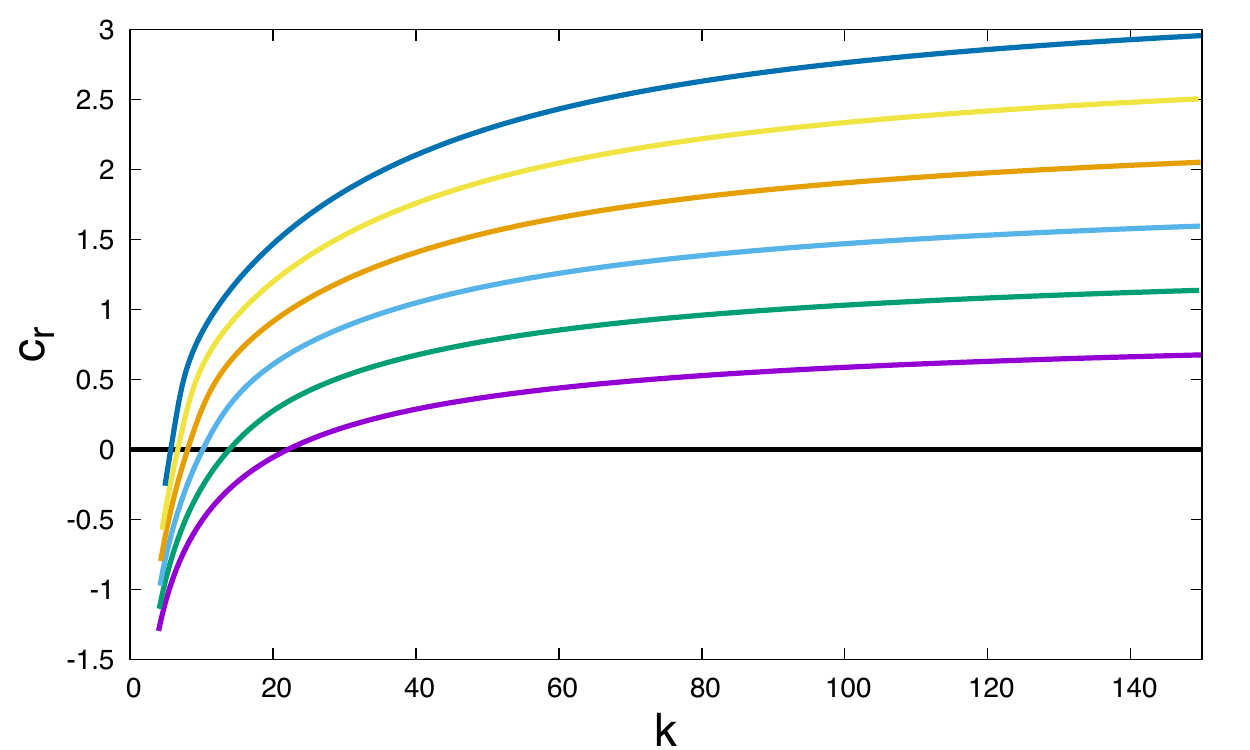}
    \caption{Color on line. Dimensional growth rates of gravity waves (top) and phase velocities (bottom), against the wavenumber. For $\delta=20\ m^{-1}$
    and $u_{0s}$ increases from $1\ m.s^{-1}$ to $3.5\ m.s^{-1}$ by increment of $0.5\ m.s^{-1}$ (the Froude number varies from 1.43 to 5). The lower curves (in the top and bottom figures) correspond to $u_{0s}=1\ m.s^{-1}$.}
    \label{fig:kcicr_delta_20}
\end{figure}

To the best of our knowledge we are not aware of detailed investigations on the linear stability of the exponential current to gravity wave perturbations ($We=\infty$).
\newline
Rippling instabilities are generated at the surface of shear layers of few millimeters whereas, for instance, the length scale of the characteristic thickness of shear layers due to discharge from river estuaries or tidal currents is much larger. The characteristic shear layer thickness at the Mouth of the Columbia River is of $\mathcal{O}(18 \, m)$ and $\mathcal{O}(13.5 \, m)$ for the ebb and flood currents, respectively. Consequently, we can expect that the length scales of the unstable perturbations occurring on the surface of currents at river estuaries belong to the class of gravity waves. This oceanic current example has motivated our investigation on the linear stability of the exponential current to pure gravity wave disturbances.
\vspace{0.15cm}
\newline
The dimensionless equation (\ref{analytic_neutral_curve}) for gravity waves reads
\begin{equation}
    \sqrt{1+K^2}-1-\frac{1}{Fr^2}=0,
    \label{dimensionless_marginal_equation_gravity}
\end{equation}
The analytic expression of the marginal curve in the ($K,Fr$) plane is
\begin{equation}
    Fr=\frac{1}{\sqrt{\sqrt{1+K^2}-1}}
    \label{dimensionless_marginal_equation_K_Fr},
\end{equation}
and the dimensionless marginal wavenumber is
\begin{equation}
    K_{marginal}=\sqrt{\frac{2}{Fr^2}+\frac{1}{Fr^4}}.
    \label{Marginal_wavenumber_gravity}
\end{equation}
In dimensional form the marginal wavenumber reads
\begin{equation}
    k_{marginal}=\sqrt{\frac{2g\delta}{u_{0s}^2}+\frac{g}{u_{0s}^4}}.
    \label{Marginal_dimensional_wavenumber_gravity}
\end{equation}
\begin{figure}
    \centering
    \includegraphics[width=0.9\linewidth]{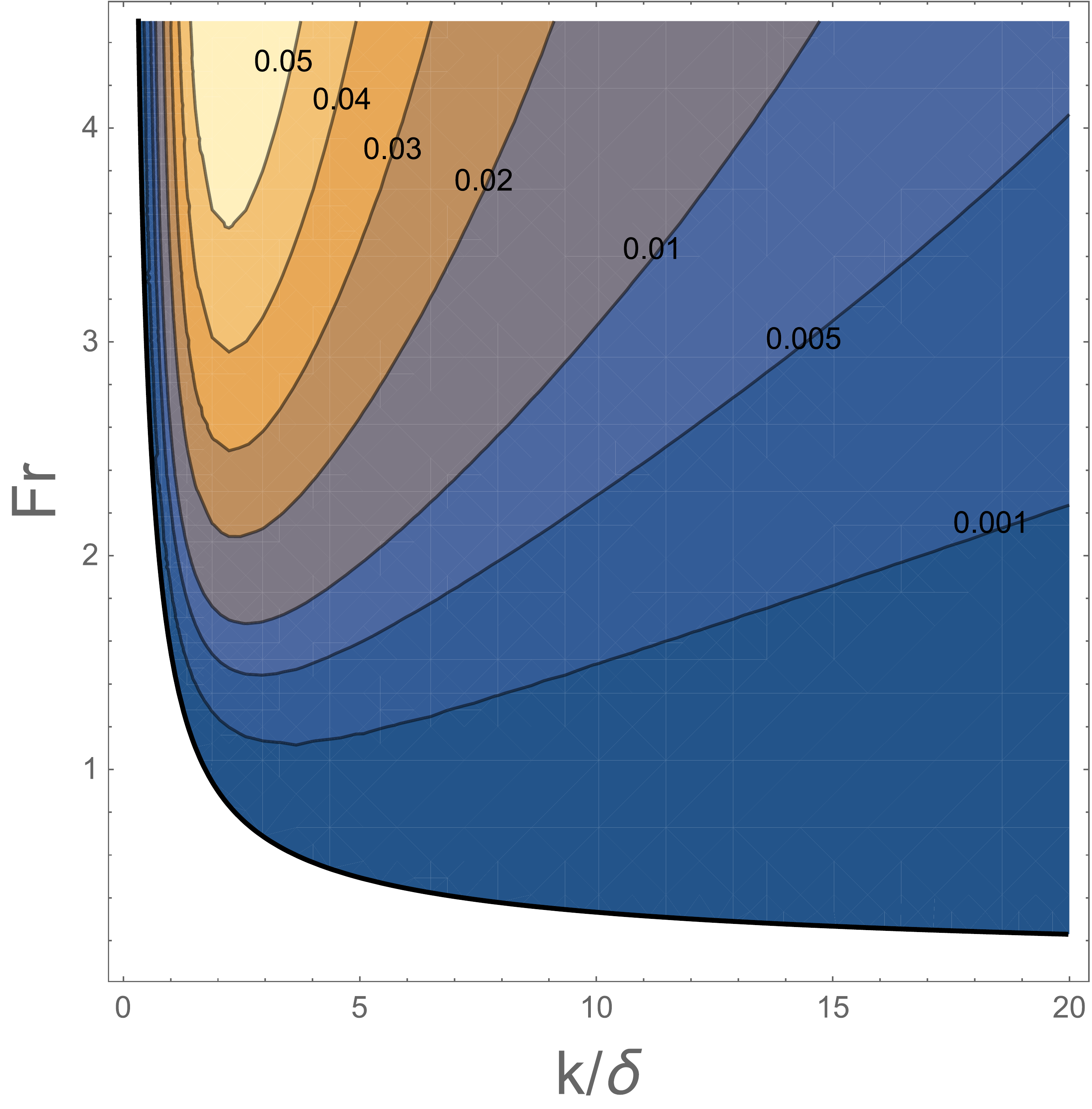}
    \caption{Color on line. Contour lines of dimensionless growth rates, $k c_i/(u_{0s}\delta)$, plotted in the $(k/\delta, Fr )$ plane. The thick black solid line corresponds to the marginal curve given by equation (\ref{dimensionless_marginal_equation_K_Fr}).
    }
    \label{fig:Stability_Fr_k/delta}
\end{figure}
\begin{figure}
    \centering
    \includegraphics[width=.48\linewidth]{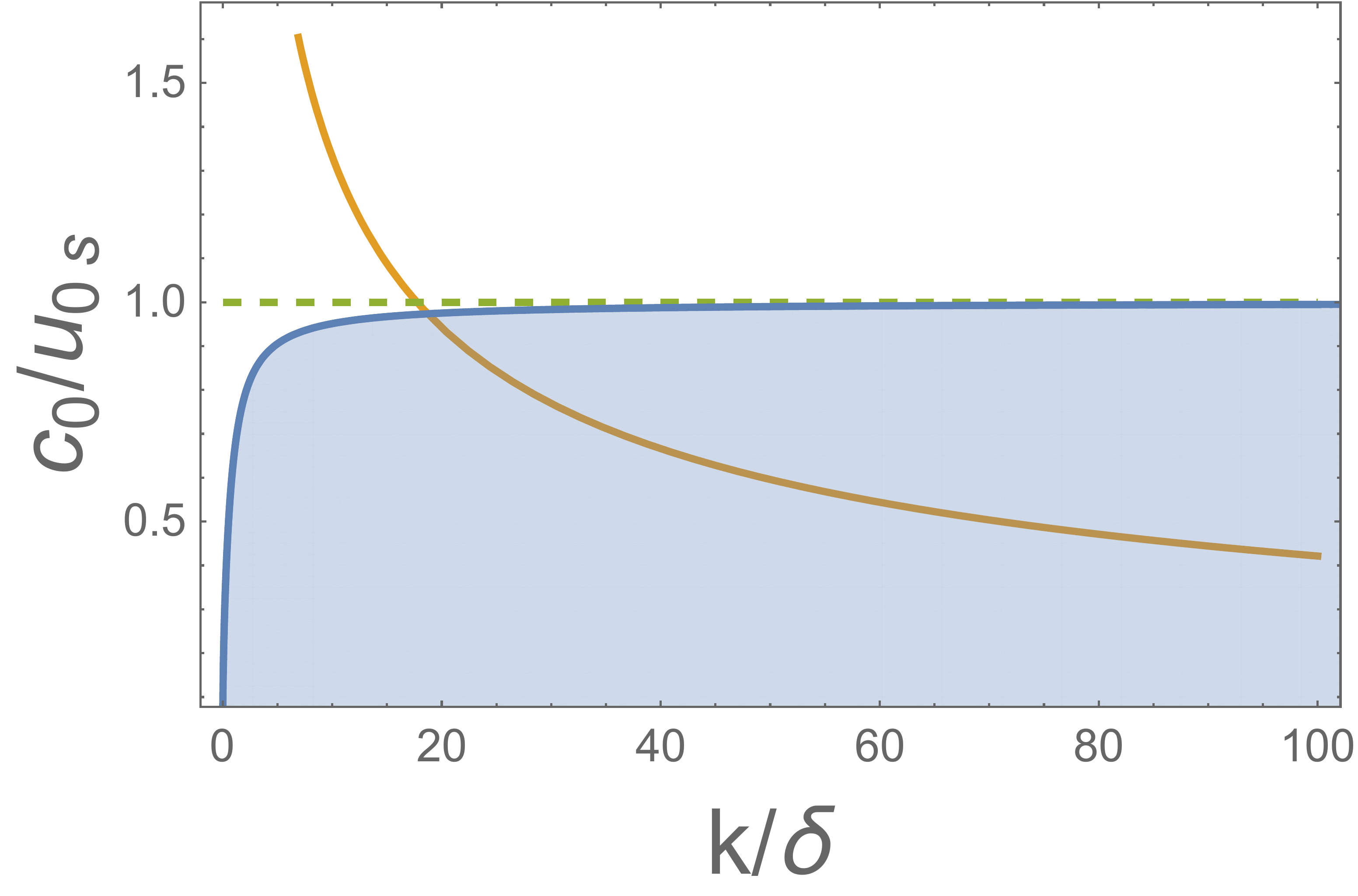}
    \includegraphics[width=.48\linewidth]{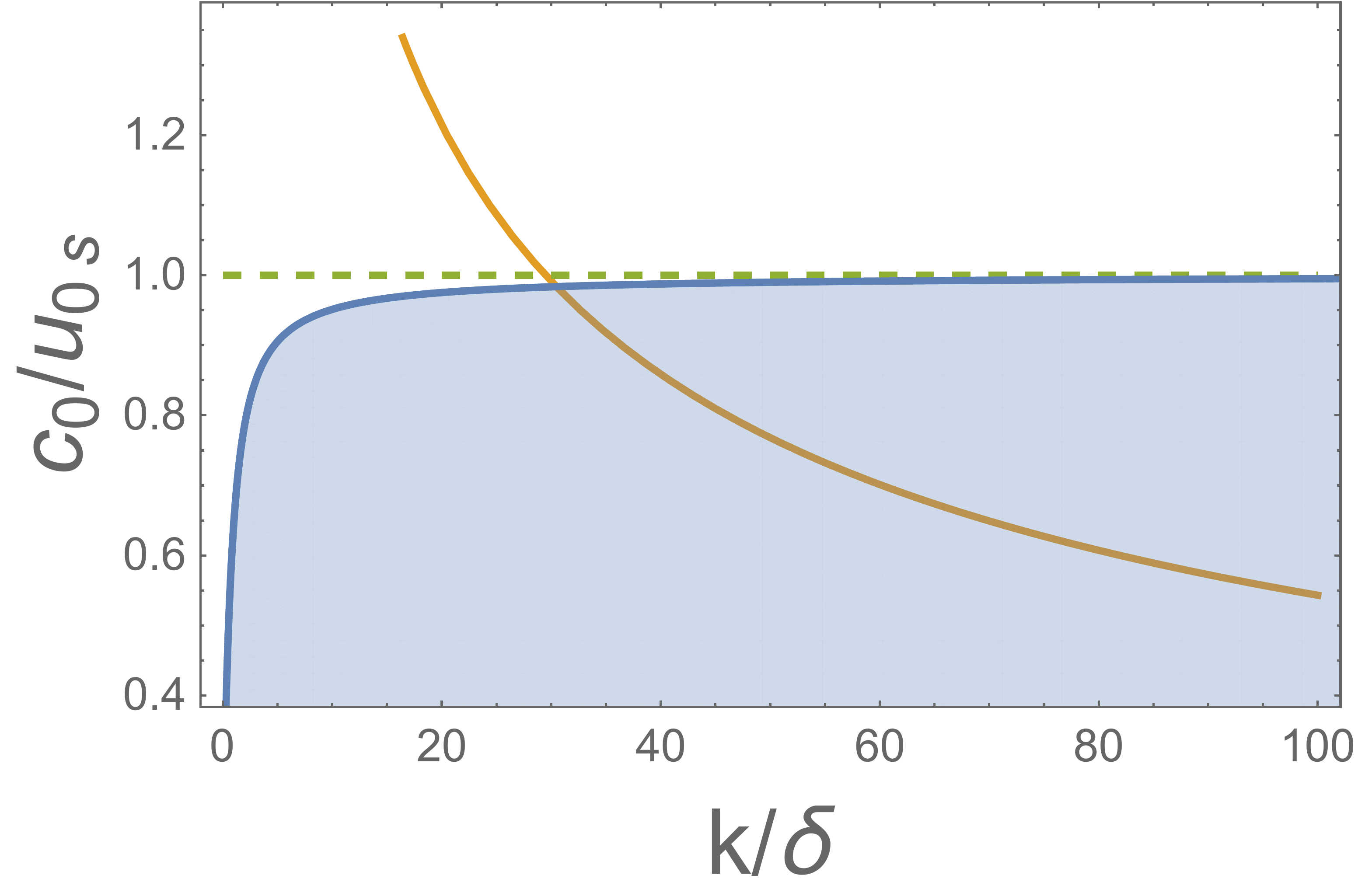}
    \caption{Color online. Stability diagram of current velocity profiles measured at the Mouth of the Columbia River. Left: Ebb current with $(u_{0s},\delta)=(-3.17\, m.s^{-1}, 0.055 \, m^{-1})$. Right: Flood current with $(u_{0s},\delta)=(2.12\, m.s^{-1}, 0.074 \, m^{-1})$. The marginal wavenumber is defined by the intersection of the marginal curve (in blue) with the graph of the linear dispersion relation (in orange)}. 
    \label{fig:Stability_c0/uos_k/delta_Ebb_Flood}
\end{figure}
\begin{figure}
    \centering
    \includegraphics[width=0.9\linewidth]{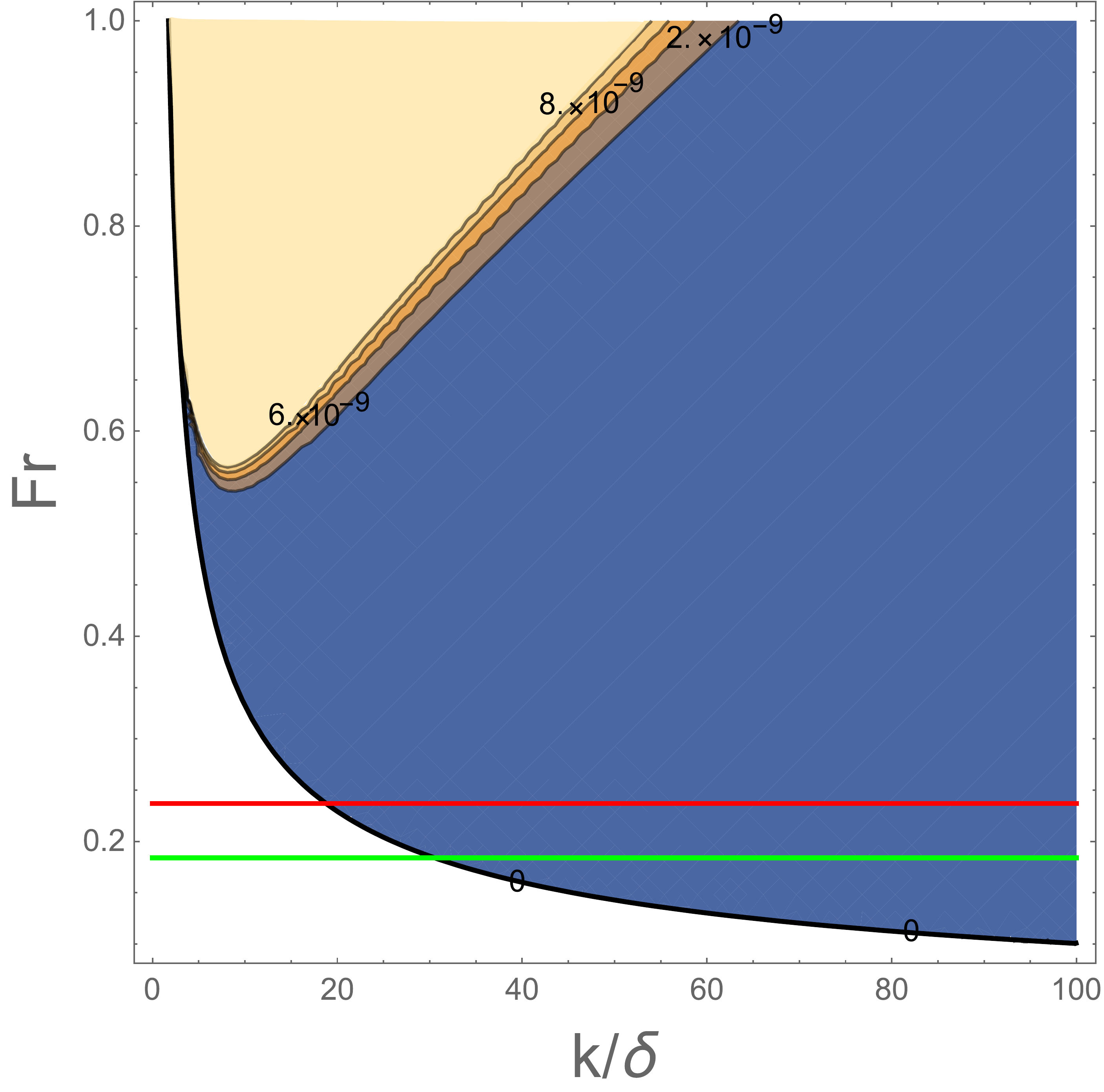}
    \caption{Color on line. Stability of ebb and flood currents at the Mouth of the Columbia River in the $(k/\delta, Fr )$ plane with some contour lines of dimensionless growth rates, $k c_i/(u_{0s}\delta)$. The red and green horizontal lines correspond to $Fr=0.237$ (ebb current) and $Fr=0.184$ (flood current), respectively. Ebb and flood currents are unstable but their growth rates are less than $10^{-9}$.}
    \label{fig:Stability_Fr_k/delta_Ebb_Flood}
\end{figure}
Figure \ref{fig:Stability_Fr_k/delta} shows the contour lines corresponding to different values of the dimensionless growth rate plotted in the ($k/\delta, Fr$) plane. One can observe that the dimensionless growth rate decreases as the Froude number decreases. 
Figure \ref{fig:Stability_c0/uos_k/delta_Ebb_Flood} shows the stability diagrams, in the ($k/\delta, c_0/u_{0s}$) plane, corresponding to the ebb and flood currents at the Mouth of the Columbia River. The ebb and flood currents are unstable for $k/\delta>18.72$ and $k/\delta>30.5$, respectively. Nevertheless, their growth rates are extremely weak as shown in figure \ref{fig:Stability_Fr_k/delta_Ebb_Flood}. To find more unstable exponential currents we have to consider higher values of the Froude number. For example, we have considered two series of values of the characteristic shear layer thickness and surface velocity. For $\delta=10\ m^{-1}$ and $\delta=20\ m^{-1}$ the surface velocity $u_{0s}$ increases by increment of $0.5\ m.s^{-1}$ from $1\ m.s^{-1}$ to $3.5\ m.s^{-1}$. In figures \ref{fig:kcicr_delta_10} and \ref{fig:kcicr_delta_20} are plotted the curves of the dimensional growth rate and dimensional phase velocity as a function of the wavenumber for several values of $\delta$ and $u_{0s}$. The dimensional growth rate increases as the surface velocity increases for fixed $\delta$ and its increases with $\delta$ for fixed $u_{0s}$. Note that $k_{\mathrm{max}}$ corresponding to the dimensional growth rate maximum is roughly independent of the surface velocity in the vicinity of $k=\delta$. We can conclude that the wavelength of the most unstable mode is approximately $2\pi/\delta$. The vanishing of the phase velocity $c_r$ corresponds to the marginal wavenumber values which decrease as $u_{0s}$ increases. The curves of the phase velocity present asymptotes defined by $\mathrm{lim} \ c_r$ as $k \rightarrow \infty < u_{0s}$.
\vspace{0.10cm}
\newline
In this subsection we have focused our investigation on the linear stability of the exponential current to pure gravity wave perturbations. We found that (i) the dimensionless rate of growth increases as the Froude number of the current increases (ii) thinner the shear layer is, larger the growth rate is for fixed surface velocity and (iii) the dimensional wavenumber corresponding to the most unstable perturbation is close to the inverse of the characteristic thickness of the shear layer.  
\section{Conclusion and perspective}
Currents in the upper ocean exist at very different vertical spatial scales. Wind action at the sea surface generates underlying shear currents of few millimeter thickness whereas tidal currents or currents due to river discharge present shear layer thicknesses much larger. \cite{Morland1991} and \cite{Young2014} investigated the stability of exponential currents of very thin shear layers and consequently, restricted their studies to rippling instabilities. To extend their studies we have investigated the stability of much thicker  exponential shear currents to infinitesimal gravity wave perturbations. We found that (i) the dimensionless growth rate increases with the Froude number based on the characteristic shear layer thickness and the surface velocity and (ii) the dimensional wavelength of the most unstable mode is of order of the characteristic shear layer thickness. Besides, within the framework of gravity-capillary instabilities (i) we derived an analytic expression of the critical characteristic thickness of the shear (ii) we provided a sufficient condition based on the Weber number based on the characteristic shear layer thickness and the surface velocity for the stability of the exponential current (iii) we have considered two stability diagrams in finite depth demonstrating that the size of stable domains increases as the characteristic thickness of the shear layer decreases and (iv) we have considered the stability of a thin film of liquid in an exponential shearing flow.
\vspace{0.1cm}
\newline
The next step is to compute nonlinear progressive water waves of permanent form on a stable exponential current. Consequently, it is crucial to check firstly the stability of the underlying current.

\clearpage
\newpage
\appendix
\section{Numerical method for the Rayleigh equation in finite depth}

\begin{figure}
\begin{centering}
\includegraphics[scale=1.2]{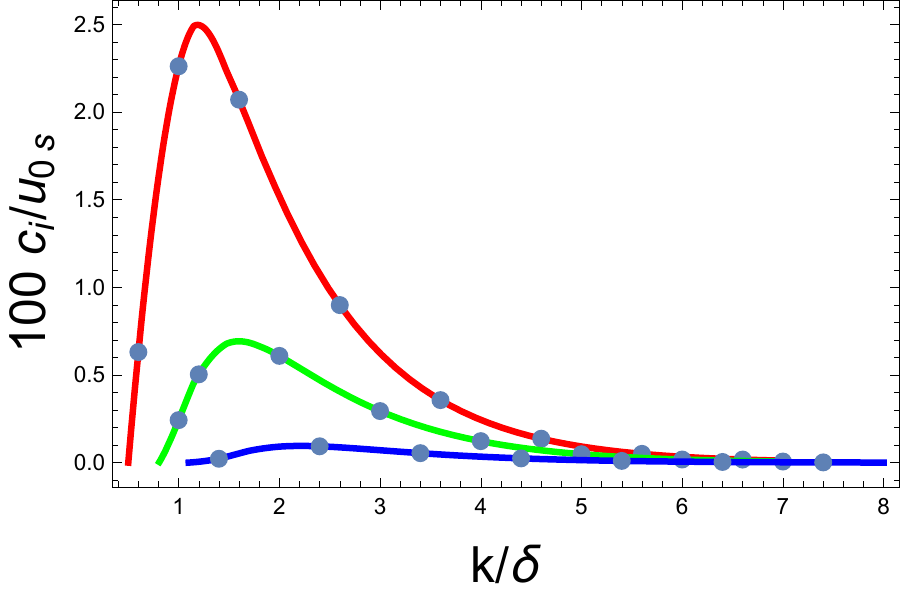}
\par\end{centering}
\caption{Color online. Validation of the numerical method using the results
of \cite{Young2014} for $W_{e}=100$ corresponding to their figure 4(e) in deep water for $Fr=2.83$ (red), $Fr=2$ (green) and $Fr=1.41$ (blue). Dots correspond to finite depth results obtained numerically using our method with $h=2.5 \lambda_m$. \label{fig:ValidationYW}}
\end{figure}

A Newton method is used to solve the Rayleigh equation with boundary conditions 
 (\ref{perturbation-kinematic+dynamic-boundary-conditions}) and (\ref{bottom_condition_finite_depth}) and the dimensionless boundary conditions $v_1(0)=1$ for the unknowns $c$ and $v_{1_{y}}(0)$. At each iteration, the
Rayleigh equation is integrated (with the appropriate boundary conditions) as described in \cite{Abid1993}, \cite{Drazin1981} and \cite{Conte1959}. The present numerical method is validated using comparisons with the infinite depth results of \cite{Morland1991} by increasing $h/\lambda_{m}$ in our code. Here $\lambda_{m}$
is the wavelength of the slowest capillary-gravity wave in calm water. The results of the validation are presented in table \ref{tab:Validation-num-meth}
and a good agreement is obtained as shown in the last line of the
table. The numerical method was also validated using the results of
\cite{Young2014} in infinite depth. The validation is presented
in figure \ref{fig:ValidationYW}. In this figure solid lines correspond
to the results of \cite{Young2014}. The dots correspond to numerical results derived from our code in finite depth. Nevertheless, note that $h=2.5 \lambda_m$ corresponds in fact to deep water as shown in figure \ref{fig:growth_thin_film}. Herein, a good agreement is obtained, too.

\begin{table}
\begin{centering}
\begin{tabular}{ccc}
$c_{r}/c_{m}$ & $\sigma_{i}\lambda_{m}/c_{m}$ & $h/\lambda_{m}$\tabularnewline
$0.333312$ & $0.0425762$ & $0.57$\tabularnewline
$0.3334$ & $0.0433171$ & $1.15$\tabularnewline
$0.3334$ & $0.0433172$ & $1.72$\tabularnewline
$0.3334$ & $0.0433172$ & $2.30$\tabularnewline
$0.333$ & $0.043$ & $\infty$ (Morland \emph{et al}.)\tabularnewline
\end{tabular} 
\par\end{centering}
\caption{Validation of the numerical method. The validation is done for many
values of $u_{0s}$ and $\delta$. Only the case with $u_{0s}/c_{m}=2$
and $\delta\lambda_{m}=7$ is shown. Our results are compared with
those of \cite{Morland1991}, in deep water, by increasing
$h/\lambda_{m}$ in our code. A good agreement is found between our
numerical results and those of \cite{Morland1991} as shown in the
last line of the table. In the table, $c_{r}$ is the absolute
phase speed and $\sigma_{i}=kc_i$ is the  growth rate. Herein, $\lambda_{m}$
is the wavelength of the slowest capillary-gravity wave in calm water
and $c_{m}$ its phase speed. \label{tab:Validation-num-meth}}
\end{table}

\section{The stability of a thin inviscid film of liquid in an exponential shearing flow}

\begin{figure}
    \centering
    \includegraphics[width=12cm]{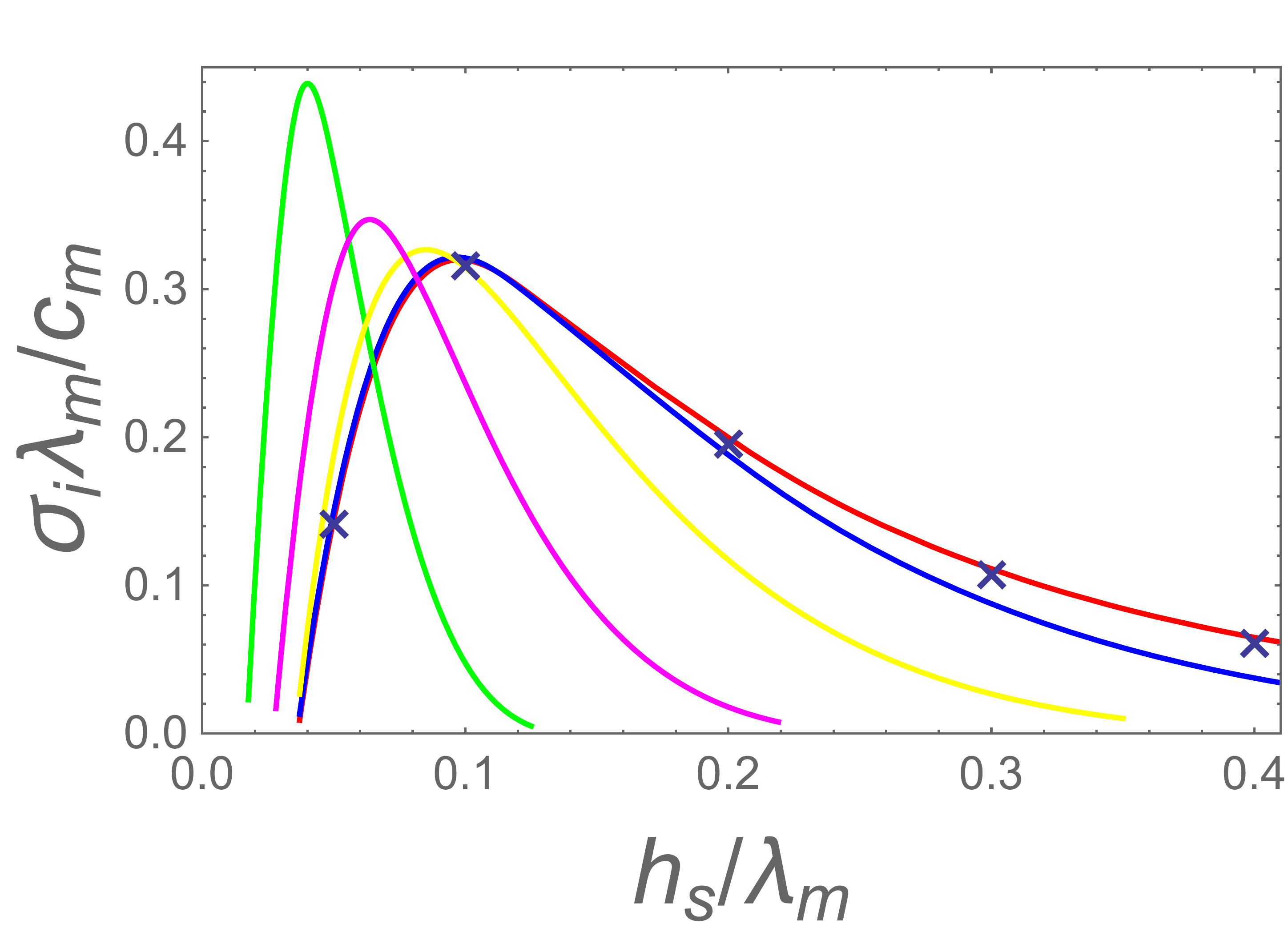}
    \caption{Color online. Dimensionless growth rate of a thin film of inviscid liquid in an exponential shearing flow, for several depths. Here, $\sigma_i=k c_i$, $h_s=1/\delta$, $c_m$ is the minimum intrinsic phase velocity of a gravity-capillary wave and $\lambda_m$ its corresponding wavelength. The surface velocity is $u_{0s}=2.5\ c_m$. The depths are (in units of $\lambda_m$): $h=1.73$ (red), $0.57$ (blue), $0.35$ (yellow), $0.22$ magenta and $0.125$ (green). Crosses correspond to infinite depth results obtained with hypergeometric functions.}
    \label{fig:growth_thin_film}
\end{figure}

\cite{Miles1960} considered the stability of a thin film of inviscid liquid in a linear shear current. He inferred that $We=\rho u_{0s}^2 h/\sigma < 3$ is a sufficient condition for stability. The fact that there
can be no energy transfer between an inviscid shear flow and a travelling wave disturbance, in the absence of profile curvature, prevented Miles to draw a conclusion concerning the instability for a linear shear, even for waves with a phase velocity in the range of the shear velocity profile. Within the framework of an exponential current, there is a profile curvature. Therefore, we have studied the instability of a thin liquid film in an exponential shearing flow. The results are presented in figure \ref{fig:growth_thin_film}. It is clear that (i) the dimensionless growth rate increases as the depth decreases and (ii) the bandwidth of the characteristic shear layer thickness corresponding to instability decreases as the depth decreases (iii) for depths greater than $1.73 \ \lambda_m$, growth rates are like those obtained in an infinite depth (when the surface velocity is equal to $2.5 \ c_m$).

\newpage
\bibliographystyle{jfm}
%
%

%
%
%
\newpage 

\backsection[Supplementary data]{No supplementary data.}

\backsection[Acknowledgements]{}

\backsection[Funding]{This research received no specific grant from any funding agency, commercial or not-for-profit sectors.}

\backsection[Declaration of interests]{The authors report no conflict of interest.}

\backsection[Data availability statement]{The data that support the findings of this study are available  upon request.}

\backsection[Author ORCIDs]{C. Kharif, https://orcid.org/0000-0003-0716-8183; M. Abid, https://orcid.org/0000-0002-0438-4182}

\backsection[Author contributions]{Malek Abid: Formal analysis; Investigation; Project administration; Software; Validation; Visualization; Writing; Data curation. Christian Kharif: Formal analysis; Investigation; Project administration; Methodology; Validation; Writing.}

\end{document}